\documentclass[aps, pra, twocolumn, groupedaddress, superscriptaddress, floatfix, longbibliography]{revtex4-1}

\usepackage{amsmath}
\usepackage{amssymb}
\usepackage{graphicx}
\usepackage{float}
\usepackage[english]{babel}
\usepackage{dsfont}
\usepackage{epstopdf}
\usepackage{soul}
\usepackage{color}
\usepackage{braket}
\usepackage[normalem]{ulem}
\usepackage[colorlinks=true,
            linkcolor=blue,
            urlcolor=blue,
            citecolor=blue]{hyperref}

\usepackage{lineno}
\definecolor{SMblue}{rgb}{0,0.0, 0.0} 
\definecolor{SMblue2}{rgb}{1,0.3,0}
\definecolor{scolor}{rgb}{0.1,0.1,1}%
\definecolor{RHG}{rgb}{0,0.5,0}
\definecolor{tc}{rgb}{0,0.4,0.8}            
\newcommand{\avg}[1]{\langle #1 \rangle}

\newcommand{\creop}[1]{\hat{a}^{\dagger}_{#1}}
\newcommand{\aniop}[1]{\hat{a}_{#1}}

 \emergencystretch=\maxdimen
\begin{document}


\title{Phase and Photon Number Dependent NOON State Localization\\ in Flat Band Lattices}

\author{Rishav Hui}
\affiliation{Department of Physics, Indian Institute of Science, Bangalore 560012, India}
\author{Trideb Shit}
\affiliation{Department of Physics, Indian Institute of Science, Bangalore 560012, India}
\author{Marco Di Liberto}
\affiliation{Dipartimento di Fisica e Astronomia ``G. Galilei" \& Padua Quantum Technologies Research Center, Universit\`a degli Studi di Padova, I-35131 Padova, Italy}
\affiliation{Istituto Nazionale di Fisica Nucleare (INFN), Sezione di Padova, I-35131 Padova, Italy}
\author{Diptiman Sen}
\affiliation{Centre for High Energy Physics, Indian Institute of Science, Bangalore 560012, India}
\author{Sebabrata Mukherjee}
\email{mukherjee@iisc.ac.in}
\affiliation{Department of Physics, Indian Institute of Science, Bangalore 560012, India}

\begin{abstract} 
Flat-band lattices supporting compact localized states provide a versatile platform for exploring unconventional transport phenomena in photonic, ultracold atomic, electronic, and other systems. Here, we investigate the transport of path-entangled multi-photon NOON states in a flat-band rhombic lattice and observe intriguing localization-delocalization features that depend on both the phase and photon number of the NOON states. To experimentally emulate photon number correlations, we develop an intensity correlation measurement protocol using coherent laser light with tunable relative phases. We first apply this protocol to show spatial bunching and anti-bunching of two-photon NOON states in a one-dimensional waveguide lattice. In the rhombic lattice, we show that for an even (odd) photon number $N$, localization occurs at $0 \, (\pi)$ phase of the NOON state with a probability of $2^{1-N}$, which is demonstrated up to eight photons. Our results open an exciting route towards understanding the dynamics of correlated photons in complex photonic networks.
\end{abstract}
\maketitle

\section{Introduction}\label{intro}

Certain lattice configurations support perfectly non-dispersive (flat) bands~\cite{leykam2018artificial, mukherjee2015observation, vicencio2015observation, taie2015coherent, baboux2016bosonic, slot2017experimental, chase2024compact, guzman2014experimental}, resulting in intriguing localization effects in the absence of disorder and interactions.
Flat bands have been explored in various contexts, including unusual ferromagnetic ground states~\cite{tasaki2008hubbard}, 
magnetic-field-induced Aharonov-Bohm caging~\cite{vidal1998aharonov, longhi2014aharonov, mukherjee2018experimental, martinez2023flat, chen2025interaction}, inverse Anderson transition~\cite{goda2006inverse, li2022aharonov, zhang2023non}, superfluidity~\cite{peotta2015superfluidity}, and unconventional superconductivity~\cite{cao2018unconventional}.
Flat-band localization of optical states has been primarily studied using classical light waves~\cite{mukherjee2015observation, vicencio2015observation, mukherjee2015observationRhombic, xia2016demonstration}. It is of great interest to understand how multi-photon quantum states evolve within such flat-band lattices, and how the interplay between band structure and non-classical initial states influences quantum interference.
Photonic platforms provide a natural playground where the transport
of quantum states of light can reveal various phenomena, such as correlated quantum walks~\cite{peruzzo2010quantum, sansoni2012two, poulios2014quantum}, boson sampling~\cite{broome2013photonic, tillmann2013experimental}, Bloch oscillations~\cite{bromberg2010bloch, lebugle2015experimental} and Anderson localization~\cite{crespi2013anderson} of entangled photons. Specifically, waveguide networks offer a scalable and flexible platform for the discovery of new fundamental science~\cite{christodoulides2003discretizing, garanovich2012light, schwartz2007transport, lahini2008anderson,
rechtsman2013photonic, hafezi2013imaging}, as well as for the development of practical quantum technologies~\cite{o2009photonic, flamini2018photonic}.

The combined task of generating quantum states with a large number of highly entangled photons (e.g., NOON states), controlling their transport in multi-port coupled photonic circuits while maintaining coherence, and performing their high-fidelity detection constitutes a substantial experimental challenge~\cite{Afek2010, pan2012multiphoton, matthews2009manipulation, matthews2011heralding, los2024high, shih2003entangled}. 
Specifically, as the number of photons 
$N$ increases, NOON states
become highly prone to loss, which degrades the entanglement and quantum coherence~\cite{rubin2007loss, kacprowicz2010experimental}.
In this context, carefully designed photonic simulators are useful for predicting quantum correlations in complex photonic networks~\cite{bromberg2009quantum, keil2010photon, shit2024probing, Keil2011biphoton, mikhalychev2022emulation}. Indeed, using a mathematical mapping, quantum correlations of two indistinguishable particles have been experimentally simulated~\cite{keil2010photon, shit2024probing} by measuring two-point intensity correlations~\cite{hanbury1956test}. In this work, we propose a {\it generalized} intensity correlation measurement protocol for emulating the evolution of $N$-photon NOON states $\psi_{m, m'}(N, \alpha)\!=\!\frac{1}{\sqrt{2N!}}(a_m^{\dagger N}+e^{-i\alpha}a_{m'}^{\dagger N}) \ket{0}$, initially coupled to the $m$-th and $m'$-th sites of laser-fabricated~\cite{davis1996writing, szameit2010discrete} photonic lattices. To demonstrate the effectiveness of this protocol, we first consider two-photon NOON states coupled to the neighboring sites of a one-dimensional lattice, which exhibit spatial bunching and anti-bunching for $\alpha\!=\!\pi$ and $0$, respectively~\cite{hong1987measurement, matthews2013observing}.
Interestingly, for a flat-band rhombic lattice, we show that the localization and delocalization of photon number correlations depend on both the phase and the photon number of the initial state, which is coupled at the upper and lower sites of a unit cell. The NOON state exhibits a highly nontrivial localization behavior depending on the parity of the total photon number $N$. Specifically, when $N$ is even (odd), all photons occupy the flat band at $\alpha\!=\!0 \, (\pi)$, with a probability of $2^{1-N}$.  
For the opposite phases, i.e., $\alpha\!=\!\pi \, (0)$, complete localization is not observed, as the probability of all photons in the flat-band is zero. 
The experiments conducted up to $N\!=\!8$ demonstrate good agreement with theoretical prediction.

The paper is organized as follows. In Sec.~\ref{model}, we briefly review the photon-number correlations associated with path-entangled NOON states. 
We then introduce the intensity-correlation protocol and establish its mapping to the photon-number correlation. 
In Sec.~\ref{Bunch_AntiBunch}, we apply the protocol to 
experimentally construct photon-number correlations and demonstrate bunching and anti-bunching in a one-dimensional waveguide lattice. 
In Sec.~\ref{FB_rhombic}, we consider a flat-band rhombic lattice and demonstrate the central result of phase- and photon-number-dependent localization of NOON states up to $N\!=\!8$ photons. Finally, in Sec.~\ref{conclusions}, we summarize the main findings and outline possible future directions.\\

\begin{figure*}[htb]
\centering
\includegraphics[width=1.00\linewidth]{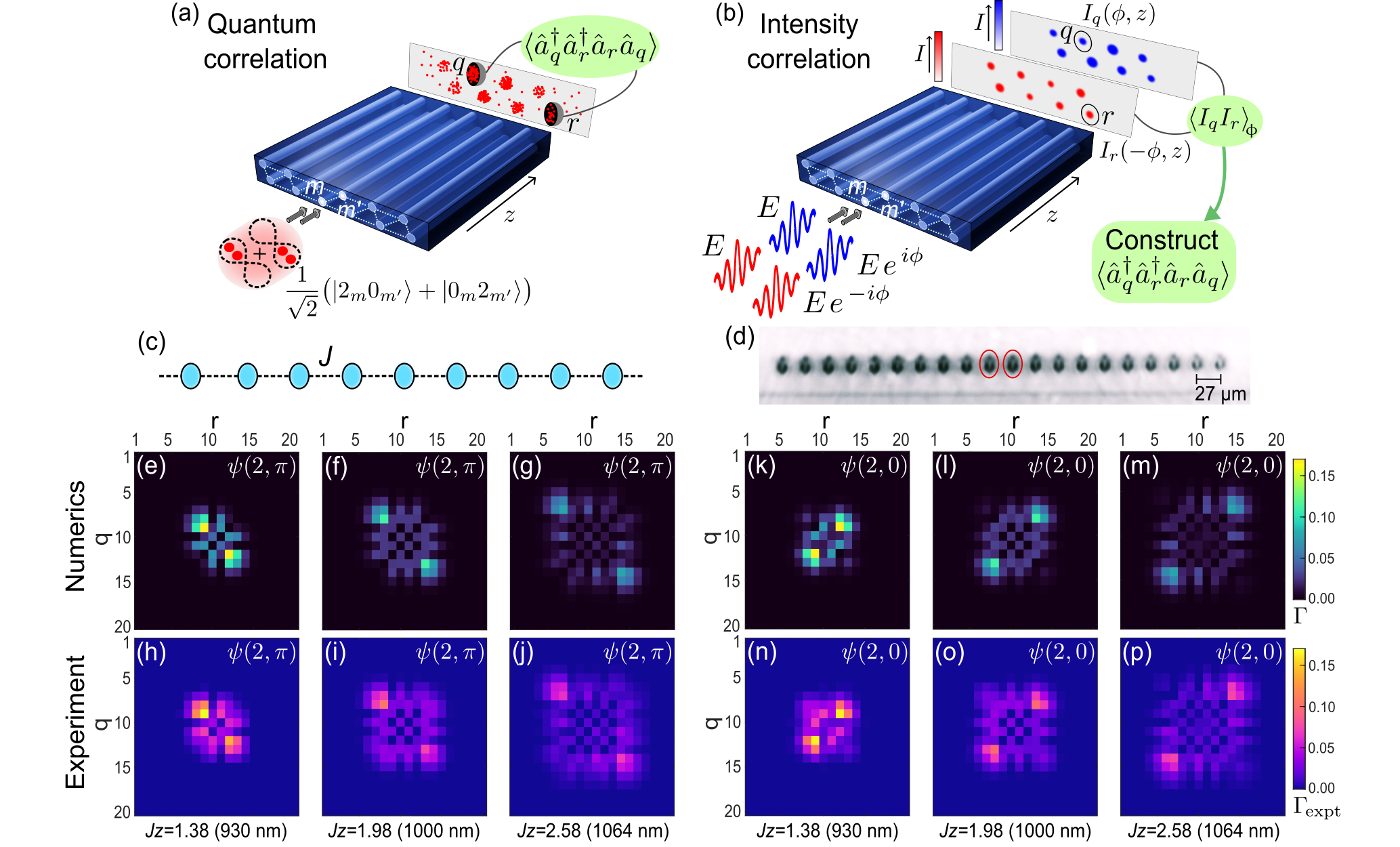}
    \caption{(a, b) Simplified schematic of quantum and intensity correlation measurement protocols for $N\!=\!2$ in a waveguide lattice.
    For the presentation purpose, here we show a zig-zag lattice. The quantum correlation matrix is constructed from intensity correlations, following Eq.~\ref{Eqn_3}.
    (c) Sketch of the one-dimensional photonic lattice used in the experiment, with nearest-neighbor coupling $J$. (d) Micrograph of a laser-fabricated photonic lattice with twenty sites. The red circles indicate $\{m,m'\}\!=\!\{10,11\}$ where coherent input states with equal intensities and desired phases are coupled.
    (e-g) Numerically calculated photon number correlations ($\Gamma$) for the two-photon NOON state $\psi_{10,11}(2, \alpha\!=\!\pi)$ showing bunching for three different effective propagation distances $Jz$. 
    (h-j) Experimentally constructed $\Gamma_{\text{expt}}$ associated with (e-g). 
    Different $Jz$ are experimentally achieved by tuning the wavelength of light (indicated below). 
    (k-m) Similar to (e-g) for $\psi_{10,11}(2, \alpha\!=\!0)$ exhibiting photon anti-bunching. 
    (n-p) Experimentally obtained $\Gamma_{\text{expt}}$ associated with (k-m). The initial states are indicated in (e-p), omitting the subscripts $\{m, m'\}$.
    }
    \label{Fig_1}
\end{figure*}


\section{Model and the proposed \\ measurement protocol}
\label{model}

Consider the propagation of photons through a periodic array of evanescently coupled optical waveguides. For a single photon initially coupled to the $j$-th site, the evolution of the bosonic creation operator is given by the Heisenberg equation~\cite{bromberg2009quantum}, $i\partial_z {\creop{j}(z)} \!=\! \sum_{j'} H_{jj'}\creop{j'}(z),$
where $z$ is the propagation distance, and $H_{jj'}$ is the element of the single-particle Hamiltonian $\hat{H}$, containing the coupling strength parameters and on-site propagation constants. Integrating the above equation, we obtain $\hat{a}_j^{\dagger}(z)\!=\! \sum_{j'} U_{jj'}(z) \hat{a}^{\dagger}_{j'}(0)$, where $U_{jj'}(z)$ is the $\{j, j'\}$-th element of the propagator $\exp{(i\hat{H}z)}$, i.e., the probability amplitude of finding the photon at the $j'$ site at a propagation distance $z$. The correlations between photons and their non-classical dynamics can be captured by photon number correlations. Considering a two-photon NOON state $\psi_{m,m'}(2, \alpha)$, the photon number correlation at the $\{q, r\}$ sites is given by,
\begin{align}
&\Gamma(q,r;m,m')\!=\!\langle \hat{a}^{\dagger}_q \hat{a}^{\dagger}_r  \hat{a}_r \hat{a}_q \rangle \nonumber \\&
= |U_{q m}(z)U_{r m}(z)+e^{i\alpha}U_{q m'}(z)U_{r m'}(z)|^2.
\label{Eqn_1}
\end{align}
The off-diagonal element of the correlation matrix represents the probability of finding one photon at the $q$-th site and its partner at the $r$-th site. The joint probability of detecting both photons at the same site $q$ is given by half of the magnitude of the $q$-th diagonal element.

To experimentally construct $\Gamma(q, r; m, m')$ in Eq.~\eqref{Eqn_1}, we consider the scalar-paraxial transport of light waves through a waveguide array. For initial states coupled to two sites, $m$ and $m'$, with equal intensity and a tunable relative phase $\phi$,  we define the following generalized spatial intensity correlation [see Fig.~\ref{Fig_1}(a, b)] at a propagation distance $z$,
\begin{align}
&G(q, r; m, m') \!=\! \avg{I_q(f_1(\phi), z)I_r(f_2(\phi), z)}_{\phi \in [0, 2\pi]}  \nonumber\\ &
= \frac{1}{2\pi}\int_{0}^{2\pi} \text{d}\phi \, I_q(f_1(\phi), z)I_r(f_2(\phi), z),
\label{Eqn_2}
\end{align}
where $\langle \cdot \rangle$ denotes the phase averaging, and $I_q(f_1(\phi), z)\!=\!\frac{1}{2}|U_{qm} \!+\! U_{qm'} e^{if_1(\phi)}|^2$ is the normalized intensity at the $q$-th site for an initial phase difference of $f_1(\phi)\! \in\! [0,2\pi]$, which is a linear function of $\phi$. Here, we consider $f_1(\phi)+f_2(\phi)\!=\!\alpha$ (see Appendix~\ref{secS_Two_photon_NOON_states}) and experimentally construct the quantum correlations matrix for $\psi_{m, m'}(2, \alpha)$ in the following way
\begin{eqnarray} \label{Eqn_3}
{\Gamma_{\text{expt}}(q,r;m,m') \!=\! 4G(q, r; m, m')-I^{q}_m I^{r}_{m'}-I^{r}_m I^{q}_{m'}}\, , \quad
\end{eqnarray}
where $I_m^q(z)$ is the intensity at the $q$-th site for the initial excitation at the $m$-th site only. 
Eq.~\eqref{Eqn_3} is an exact mathematical analog of Eq.~\eqref{Eqn_1}. All the quantities on the right-hand side of Eq.~\eqref{Eqn_3} can be obtained in a phase-averaged measurement with coherent laser light.

We further highlight that $G(q,r; m,m')$ can be mapped to the quantum correlations of two indistinguishable 
anyons~\cite{kwan2024realization}, initially located at two different lattice sites, by considering
$|f_1(\phi)\!-\!f_2(\phi)|\!=\!\alpha$; see Appendix~\ref{secS_Anyonic correlation} for details. The special cases of bosonic and fermionic correlations~\cite{bromberg2009quantum, keil2010photon, shit2024probing} are obtained for $\alpha\!=\!0$ and $\pi$, respectively.

Interestingly, by defining the three-point intensity correlation as $G(p,q,r; m,m') \! = \! \braket{I_p(f_1(\phi), z) I_q (f_2(\phi), z) I_r(f_3(\phi), z)}_{\phi\in[0,2\pi]}$, and constraining $f_1(\phi)+f_2(\phi)+f_3(\phi) \!=\! \alpha$, we construct the quantum correlations of the three-photon NOON state as
\begin{eqnarray}
&\Gamma_{\text{expt}}(p,q,r; m, m')\!=\!8G(p,q,r; m, m')- \nonumber \\& 
\big(I_{m'}^p I_m^q I_m^r + I_m^p I_{m'}^q I_{m'}^r 
+I_m^p I_{m'}^q I_m^r \nonumber \\ &
+I_{m'}^p I_m^q I_{m'}^r+I_m^p I_m^q I_{m'}^r+ I_{m'}^p I_{m'}^qI_m^r \big) \, . \,
\label{Eqn_4}
\end{eqnarray}
The above protocol can be generalized for $N$-photon NOON states, as discussed in Appendix~\ref{secS_N-photon_NOON_states}.

\begin{figure*}[htb]
    \centering
\includegraphics[width=1.00\linewidth]{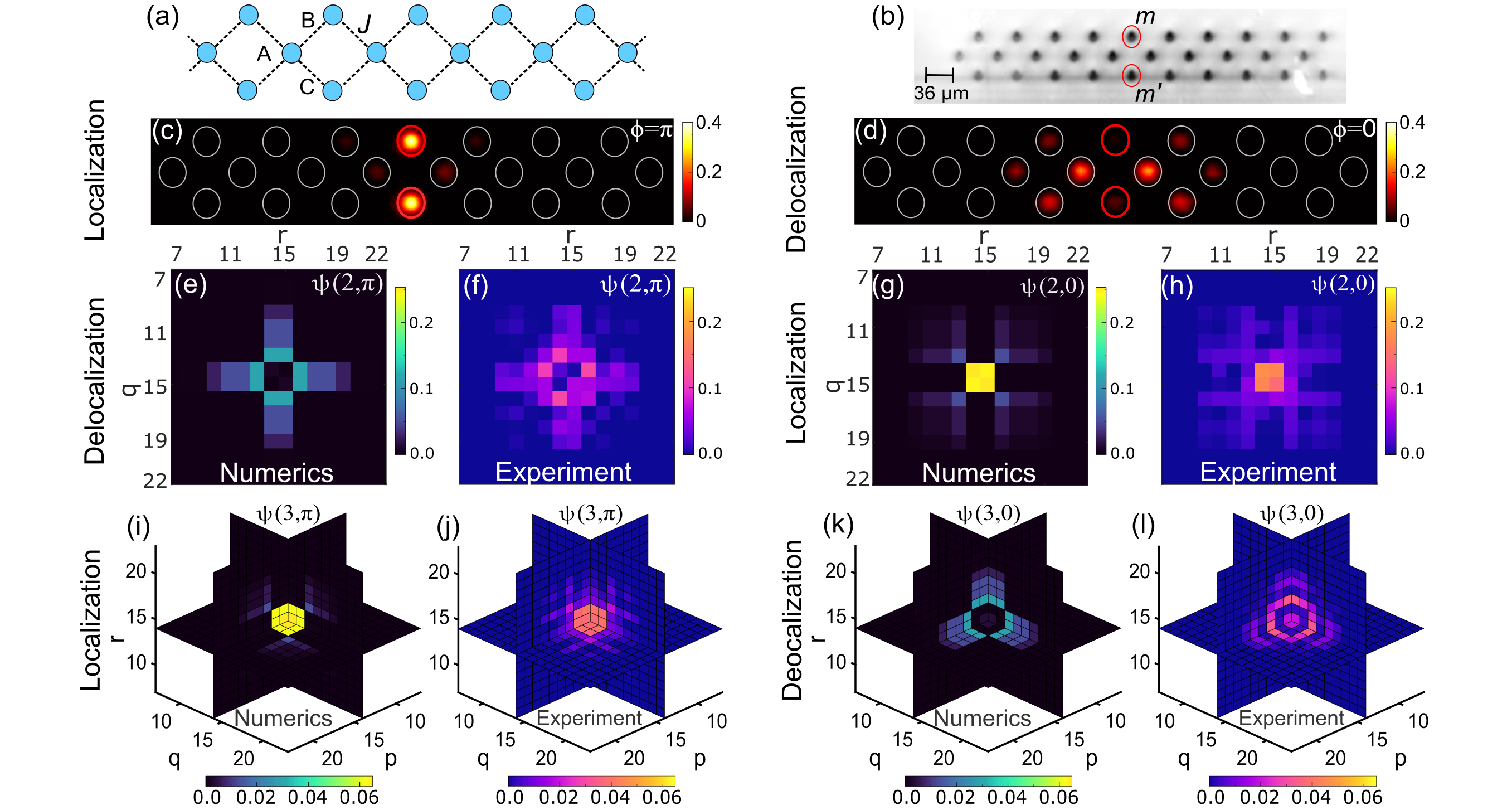}
    \caption{%
    (a) Sketch and (b) micrograph of a photonic rhombic lattice. The red circles indicate $\{m,m'\}\!=\!\{14,15\}$ where the initial state is coupled for all experiments.
    (c, d) Measured intensity patterns at $Jz\!=\!0.91$ 
    for two-site input states exciting the flat band and dispersive band of the rhombic lattice, respectively. 
    (e) Numerically obtained, and (f) experimentally constructed photon number correlations for the two-photon NOON state $\psi_{14,15}(2, \alpha \!=\! \pi)$ showing delocalization.
    (g, h) same as (e, f) for  $\psi_{14,15}(2, \alpha\!=\!0)$ showing strong localization.
    (i, j) Numerics and experiments showing localization for the third-order correlation map for a three-photon NOON state $\psi_{14,15}(3, \alpha \!=\! \pi)$, respectively. 
    (k, l) same as (i, j) for $\psi_{14,15}(3, \alpha\!=\!0)$ showing delocalization.
    The localization and delocalization depend on the phase as well as the photon number of the NOON states.
    }
    \label{Fig_2}
\end{figure*}

\vspace{12pt}

\section{Bunching and anti-bunching}\label{Bunch_AntiBunch}

To perform the intensity correlation measurement shown in Fig.~\ref{Fig_1}(b), a collimated optical beam at wavelength $\lambda$ is split into two parts and are coupled to two consecutive sites $\{m, m'\}\!=\!\{10, 11\}$ of a fs laser-fabricated one-dimensional lattice, Fig.~\ref{Fig_1}(c, d). Before coupling to the lattice, one of the beams is reflected by a spatial light modulator (SLM) to tune the relative phase with a step size of $\pi/128$. The $\{q, r\}$ element of intensity correlation at the output is then obtained by phase averaging the product of the output intensities $I_q(\phi, z)$ and $I_r(-\phi, z)$. 
We used a $z\!=\!40$-mm-long photonic lattice, and $J(\lambda)z$ was varied by tuning the wavelength of light~\cite{szameit2010discrete, sinha2025probing}; see also Appendices~\ref{secS_Fabrication details},\ref{secS_Details on intensity correlation measurement}.
In all experiments described below, the light remains confined within the bulk of the lattice during propagation; therefore, edge effects can be neglected. %

The numerically calculated photon number correlations for $\alpha\!=\!\pi$ at three different effective propagation distances $Jz$ are presented in Figs.~\ref{Fig_1}(e-g). The prominent lobes in the main diagonal of the correlation matrices indicate that the probability of joint detection of the photons is large in this case -- an effect known as {spatial} bunching. Experimentally constructed photon number correlations shown in Figs.~\ref{Fig_1}(h-j) are in excellent agreement with the numerical prediction. Photon number correlation is sensitive to the phase $\alpha$ of the NOON state -- in the case of $\alpha\!=\!0$, two photons primarily travel in opposite directions in the lattice, giving rise to anti-bunching; see prominent anti-diagonal elements in Figs.~\ref{Fig_1}(k-m) and the associated experimental results in Figs.~\ref{Fig_1}(n-p).

The dispersion of the one-dimensional lattice in momentum ($k$) space is given by $\varepsilon(k)\!=\!-2J\cos(ka)$, where $J$ is the coupling strength and $a$ is the waveguide spacing. The observed bunching and anti-bunching effects in Fig.~\ref{Fig_1} are primarily caused by the Bloch modes with maximum group velocity around $ka\!=\!\pm\pi/2$, and the phase of the NOON state determines in which direction the photons propagate. 
The NOON states $\psi_{m, m+1}(2,\alpha\!=\!0)$ and $\psi_{m, m+1}(2,\alpha\!=\!\pi)$ can be expressed in
momentum space and for $ka\!=\!\pm \pi/2$ as $\hat{a}_{\pi/2}^{\dagger} \hat{a}_{-\pi/2}^{\dagger}$ and
$(\hat a_{\pi/2}^{\dagger 2} + \hat a_{-\pi/2}^{\dagger 2})/2$, where $\hat{a}_{k}^{\dagger}$ is the photon creation operator at momentum $k$. Note that for $\alpha\!=\! 0$, the two photons travel with opposite momentum, exhibiting anti-bunching. On the other hand, the NOON state with $\alpha\!=\!\pi$ excites the $ka\!=\!\pm\pi/2$ Bloch modes such that the two photons travel together in either direction with equal probability, giving rise to the bunching effect.
In this context, we note that two indistinguishable bosons (fermions) incident on two ports of a beam splitter show bunching (anti-bunching) due to particle statistics~\cite{henny1999fermionic}. 
Whereas, the bunching and anti-bunching of the two-photon NOON state is due to the quantum interference for the specific form of the state. Interestingly, when the photon number is increased to $N\!=\!3$, NOON states in the one-dimensional lattice produce the same $\Gamma$ for $\alpha\!=\!\pi$ and $0$, see Appendix~\ref{secS_with_1D_3_photon}.

\vspace{12pt}

\section{Flat-band rhombic lattice}
\label{FB_rhombic}

Now, we consider a quasi-one-dimensional photonic rhombic lattice consisting of three sites (A, B, and C) per unit cell, Figs.~\ref{Fig_2}(a, b).  In this case, the single-particle tight-binding Hamiltonian is given by $\hat{H}\!=\!-J \sum (\hat{a}_j^{\dagger}\hat{b}_j+\hat{a}_j^{\dagger}\hat{c}_j+\hat{a}_j^{\dagger}\hat{b}_{j-1}+\hat{a}_j^{\dagger}\hat{c}_{j-1})+{\text{H.c.}}$, where $J$ is the nearest-neighbor coupling, and $j$ is the unit cell index. The spectrum of the lattice consists of a perfectly flat band $\varepsilon_{0}(k)\!=\!0$ and two dispersive bands $\varepsilon_{\pm}(k)\!=\!\pm 2J\sqrt{1+\cos(kd)}$, where $d$ is the lattice constant.  The corresponding eigenmodes are given by $~(0,~1,~-1)^T/\sqrt{2}$ and $(\pm g(k),~1,~1)^T/\sqrt{2+g^2(k)}$, where $g(k)\!=\! \big(2(1+e^{-ikd})/(1+e^{ikd})\big)^{1/2}$.
Here, the compact localized states (CLS) are spatially non-overlapping and confined to a unit cell. Evidently, the flat-band CLS  can be excited by launching light at the B and C sites of a unit cell ($\{ m, m'\} \!=\! \{14,15\}$) with equal intensity and opposite phase, causing a complete localization of the initial state, as observed in Fig.~\ref{Fig_2}(c) for $Jz\!=\!0.91$. When the light is coupled to the same sites with equal phase, which excites only the dispersive bands, the initial state spreads out symmetrically away from the initially excited sites; see Fig.~\ref{Fig_2}(d).

Interestingly, for NOON states, the localization-delocalization feature in our flat band lattice can be highly dependent on the phase as well as {\it the photon number}. The photon number correlations at $Jz\!=\!0.91$ for $\psi_{14,15}(2, \pi)$ is presented in Figs.~\ref{Fig_2}(e, f). In this case, one photon remains localized, the other one travels across the lattice, and the probability of both being localized is zero. A dramatic change in the outcome can be observed by simply tuning the phase of the NOON state to $\alpha\!=\!0$. The numerical and experimental $\Gamma$ for $\psi_{14,15}(2, 0)$ are shown in Figs.~\ref{Fig_2}(g, h), respectively. Here, the probability of both photons in the flat band is significant, causing the observed localization. As shown in Fig.~\ref{Fig_S_localization_oscillation}
, the joint correlation of two photons --
either both at site B, both at site C, or one at each site -- converges to $1/4$ at long propagation distances, leading to the localization probability of $P_L\!=\!1/2$.  On a separate note, the correlation matrix at any phase $\alpha$ can be constructed using our experimental protocol, as demonstrated in Appendix~\ref{secS_with_varying_NOON_phase}.

\begin{figure}[]
    \centering
\includegraphics[width=1\linewidth]{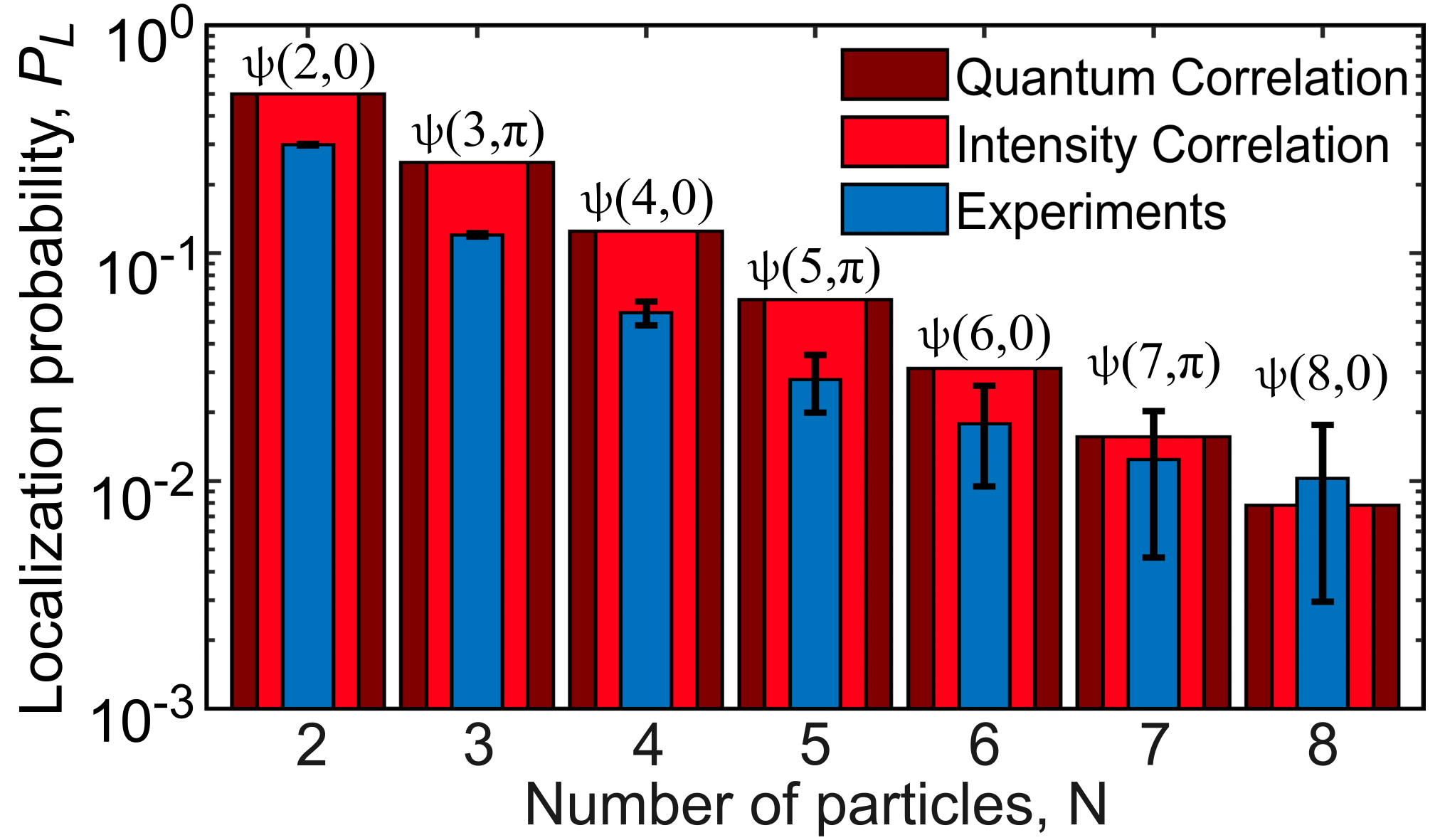}
    \caption{Localization probability of all  NOON state photons in the flat-band as a function of $N$. The initial states are indicated above each bar. The values obtained from intensity correlations (red), with a phase resolution of $\pi/128$, agree with quantum correlation calculations (brown). The blue bars show experimental results.
    }
\label{Fig_3}
\end{figure}

The above localization and delocalization of NOON state correlations flip when the photon number is increased to three. In the case of $\psi_{14,15}(3, \pi)$, the probability of detecting all three photons in the flat band is $P_L\!=\!1/4$ (see Appendix~\ref{secS_analysis of NOON state in flat-band}), resulting in the localization effect, as shown in the coordinate planes in Fig.~\ref{Fig_2}(i, j). For the $\psi_{14,15}(3, 0)$ state, there exists a nonzero probability that all three photons or some of them are delocalized; however, the probability of all three photons occupying the flat band is zero, see $\Gamma$ and $\Gamma_{\text{expt}}$ in Fig.~\ref{Fig_2}(k, l). Notice that the localization and delocalization features in Fig.~\ref{Fig_2} appear alternately with photon number for a given phase (either $0$ or $\pi$) of the NOON state. This can also be understood by expressing the initial states in the $k$-space and obtaining their contributions across different bands, as discussed in Appendix~\ref{secS_analysis of NOON state in flat-band}.

Finally, we employ the intensity correlation protocol for probing flat-band localization of NOON states with a large photon number. According to quantum correlation calculation, the probability of all photons occupying the flat band is given by $P_L(\alpha) \!=\! 2^{-N}(1 + (- 1)^N \cos(\alpha))$; see Appendix~\ref{secS_analysis of NOON state in flat-band}. In other words, for an even (odd) $N$, $P_L$ is maximum at $\alpha\!=\!0$ $(\pi)$ and goes to zero for the opposite phases, i.e., at $\alpha\!=\!\pi$ $(0)$. 
Figure~\ref{Fig_3} shows cases of maximum localization with $P_L\!=\!2^{1-N}$ as a function of $N$, alternately considering $\alpha\!=\!0$ and $\pi$. 
With our experimental step-size in controlling the relative phase $\phi$, the values of $P_L$ obtained from the intensity correlations agree with the quantum calculation up to $N\!=\!8$. 
From the experimentally constructed correlations (as in Fig.~\ref{Fig_2}), we determine the probability of finding all particles at the input sites $14$ and $15$. This probability converges to $P_L$ after some finite propagation, see Fig.~\ref{Fig_S_localization_oscillation}. The experimentally obtained $P_L$, averaged over two independent measurements, along with the standard error, is presented in Fig.~\ref{Fig_3} (blue bars).
For small $N$ values, the deviations in the experiment are primarily caused by small randomness in the lattice. For larger $N \geq 7$, another effect, i.e., the noise in $\phi$ also plays an important role.
Indeed, the accuracy of our protocol can be influenced by the  resolution and noise in the relative phase $\phi$, see Appendix~\ref{Robustness} for more details.
The results in Fig.~\ref{Fig_3} demonstrate the capability of the intensity correlation protocol for emulating NOON states with a large number of photons. 
It should be emphasized that improved resolution in $\phi$ and reduced phase noise would enable access to higher values of $N\!>\!8$.

\vspace{12pt}

\section{Conclusions}
\label{conclusions}

We have demonstrated a novel localization-delocalization effect in a flat-band rhombic lattice, with clear dependence on the phase and photon number of NOON states,  up to $N\!=\!8$. In other flat-band systems, this phenomenon can be influenced by the number of unit cells occupied by the compact localized state (CLS) and the specific input sites where the NOON state is coupled (Appendix~\ref{secS_sawtooth_lattice}). 
We note that our protocol enables the experimental construction of a quantum mechanical observable, i.e., photon-number correlation, without relying on genuine NOON states, which are experimentally challenging to realize and control for large $N$. However, this approach does not reproduce the intrinsic non-classical features of NOON states, such as entanglement-induced quantum fluctuation suppression~\cite{nagata2007beating, boto2000quantum}.

Our work opens a new avenue in the investigation of multi-particle localization, complementing other platforms such as ultracold atoms~\cite{bloch2012quantum, simon2011quantum} and Rydberg polaritons~\cite{clark2020observation}. An important open question is how such multi-particle localization-delocalization effects are influenced by disorder~\cite{bodyfelt2014flatbands},  interactions~\cite{preiss2015strongly, vidal2000interaction, vidal2002pairing}, and non-Hermiticity~\cite{owen2017dissipation, leykam2017flat}.
Our experimental protocol of constructing photon number correlations will be useful to emulate other multi-particle entangled states initially occupying more than two sites in complex photonic networks~\cite{rechtsman2013photonic, Tschernig2021, rojas2017quantum}.

\vspace{5mm}
\begin{acknowledgments}
We thank Nathan Goldman, Subroto Mukerjee, and Apoorva Patel for helpful discussions. S.M.~gratefully acknowledges support from the Indian Institute of Science (IISc) through a start-up grant; the Ministry of Education, Government of India, through the STARS program (MoE-STARS/STARS-2/2023-0716); and the Infosys Foundation, Bangalore. R.H.~and T.S.~thank IISc for their scholarships through the Integrated PhD program. 
M.D.L~acknowledges support from the  Quantum Technology Flagship project PASQuanS2, the INFN project Iniziativa Specifica IS-Quantum, and the Italian Ministry of University and Research via the Rita Levi-Montalcini program. D.S.~thanks Science and Engineering Research Board (SERB), India, for funding through Project No.~JBR/2020/000043.\\
\end{acknowledgments}

\section*{Data availability} The data supporting this study’s findings are openly available~\cite{Hui2026}.


\appendix

\setcounter{secnumdepth}{2}
\setcounter{equation}{0}
\renewcommand{\theequation}{\Alph{section}\arabic{equation}}
\setcounter{figure}{0}
\renewcommand{\thefigure}{A\arabic{figure}}%
\setcounter{section}{0}
\renewcommand{\thesection}{\Alph{section}}%

\section*{Appendix}

In this Appendix, we first show how multi-point intensity correlations obtained for specific initial states can be mathematically mapped to the photon number correlations of $N$-photon NOON states. We then discuss two-particle anyonic correlations. We provide additional experimental results for completeness. Specifically, the influence of phase resolution and noise on the accuracy of the proposed protocol is discussed. We then present two-photon NOON state correlation data in the rhombic lattice with varying $\alpha$. 
We also show three-photon correlation measurements in the one-dimensional lattice.
Next, we explain why the localization-delocalization behavior of multi-photon NOON states in the flat band rhombic lattice depends on the phase $\alpha$ of the NOON states as well as the photon number $N$. Finally, we discuss the localization-delocalization features of NOON states in a flat-band sawtooth lattice.


\section{Mapping intensity correlation to photon number correlation} \label{mapping}

\subsection{Two-photon NOON states} \label{secS_Two_photon_NOON_states}
The propagation of photons through a waveguide lattice is governed by the Heisenberg equation: $i\partial_z {\creop{j}(z)} \!=\! \sum_{j'} H_{jj'}\creop{j'}(z),$
where $z$ is the propagation distance, $H_{jj'}$ is the element of the single particle Hamiltonian $\hat{H}$, and $\creop{j}(z)$ is the bosonic creation operator at the $j$-th site of the lattice.
Integrating the above equation, we obtain
$\hat{a}_j^{\dagger}(z)\!=\! \sum_{j'} U_{jj'}(z) \hat{a}^{\dagger}_{j'}(0)$, where $U_{jj'}(z)$ is the $\{j, j'\}$-th element of the matrix $\exp{(i\hat{H}z)}$.

The photon number correlation for the two-photon NOON state $\ket{\psi_{mm'}(2,\alpha)} \!=\! {(1/2)}(\creop{m}{}^2 +e^{-i\alpha}\creop{m'}{}^2)\ket{0}$, after a propagation distance $z$, is given  by ~\cite{bromberg2009quantum}
\begin{align}
&\Gamma(q,r;m,m') \nonumber \\
& \!=\! \bra{\psi_{mm'}(2,\alpha)}\creop{q}(z)\creop{r}(z)\aniop{r}(z)\aniop{q}(z)\ket{\psi_{mm'}(2,\alpha)} \nonumber \\
&\!=\!|U_{qm}(z)U_{rm}(z)+e^{i\alpha}U_{qm'}(z)U_{rm'}(z)|^2 \, ,
 \label{Eqn_S1}
\end{align}
where we have used the bosonic commutation relations, $[\hat{a_i},\hat{a_j}^{\dagger}]\!=\!\delta_{ij}$, $[\hat{a_i},\hat{a_j}]\!=\!0$ and $[\hat{a_i}^{\dagger},\hat{a_j}^{\dagger}]\!=\!0$.

We now show how the quantum correlation of the two-photon NOON state in Eq.~\eqref{Eqn_S1} can be constructed from the two-point intensity correlation. To this end, we consider the propagation of coherent laser light coupled to the $m$ and $m'$ sites of the photonic lattice. For an initial state with equal intensity and a relative phase of $f_1(\phi)$, the 
normalized intensity at the $q$-th site of the lattice at a propagation distance $z$ is given by $I_q(f_1(\phi),z)\!=\!
(1/2) |U_{qm}(z)+U_{qm'}(z)e^{if_1(\phi)}|^2$.
Here, we express the phase factor $f(\phi)$ as a function of the relative phase of light launched at the two input sites. In our case, this can be a linear function of $\phi$ depending on the quantum state we want to emulate. 
We now define the generalized intensity correlation, a discrete analogue of Eq.~\eqref{Eqn_2}
, as
\begin{align}
&G(q,r;m,m') \nonumber\\
&= \avg{I_q(f_1(\phi),z)I_r(f_2(\phi),z}_{\phi\in[0,2\pi]}  \nonumber\\
&=(1/4)\avg{ U_{qm}U_{rm}U^{*}_{rm}U^{*}_{qm} \!+\! U_{qm}U_{rm'}U^{*}_{rm'}U^{*}_{qm} \notag \\
&+   U_{qm'}U_{rm}U^{*}_{rm}U^{*}_{qm'} \!+\! U_{qm'}U_{rm'}U^{*}_{rm'}U^{*}_{qm'} \notag\\
&  + U_{qm}U_{rm'}U^{*}_{rm}U^{*}_{qm}e^{if_2(\phi)} \!+\! U_{qm}U_{rm}U^{*}_{rm'}U^{*}_{qm}e^{-if_2(\phi)} \notag \\
&+ U_{qm'}U_{rm'}U^{*}_{rm}U^{*}_{qm'}e^{if_2(\phi)}\!+\! U_{qm'}U_{rm}U^{*}_{rm'}U^{*}_{qm'}e^{-if_2(\phi)} \notag\\
& + U_{qm'}U_{rm}U^{*}_{rm}U^{*}_{qm} e^{if_1(\phi)} \!+\! U_{qm'}U_{rm'}U^{*}_{rm'}U^{*}_{qm}e^{if_1(\phi)} \notag \\
&+ U_{qm}U_{rm}U^{*}_{rm}U^{*}_{qm'} e^{-if_1(\phi)} \!+\! U_{qm}U_{rm'}U^{*}_{rm'}U^{*}_{qm'}e^{-if_1(\phi)}\notag\\
&+ U_{qm'}U_{rm'}U^{*}_{rm}U^{*}_{qm}e^{i(f_1(\phi)+f_2(\phi))} \notag \\
&+ U_{qm}U_{rm}U^{*}_{rm'}U^{*}_{qm'} e^{-i(f_1(\phi)+f_2(\phi))}\notag\\
& + U_{qm'}U_{rm}U^{*}_{rm'}U^{*}_{qm} e^{i(f_1(\phi)-f_2(\phi))} \notag \\
&+ U_{qm}U_{rm'}U^{*}_{rm}U^{*}_{qm'}e^{-i(f_1(\phi)-f_2(\phi))}}_{\phi}
\label{Eqn_S2}
\end{align}
Here, $\langle \cdot \rangle$ denotes phase averaging over $\phi$ from $0$ to $2\pi$.
It should be highlighted that the generalized intensity correlation gives the known Hanbury-Brown-Twiss (HBT) intensity correlation~\cite{shit2024probing} in the limit of $f_1(\phi)\!=\!f_2(\phi)\!=\!\phi$.
Comparing Eq.~\eqref{Eqn_S1} and Eq.~\eqref{Eqn_S2}, we notice two extra terms, $U_{qm}U_{rm'}U^{*}_{rm'}U^{*}_{qm}\!=\!I_{m}^{q}I_{m'}^{r}$ and $U_{qm'}U_{rm}U^{*}_{rm}U^{*}_{qm'}\!=\!I_{m'}^{q}I_{m}^{r}$, in the expression of intensity correlation. These extra terms cannot be omitted by phase-averaging; however, they can be measured experimentally and then subtracted from the intensity correlation [Eq.~\eqref{Eqn_S8}]. Additionally, we note that the following relationship among $f_1(\phi)$, $f_2(\phi)$, and $\alpha$ must be satisfied to construct the quantum correlation $\Gamma$ from the intensity correlation.
\begin{align}
&\frac{1}{2\pi}\int_{0}^{2\pi}e^{i(f_1(\phi)+f_2(\phi))}d\phi \!=\! e^{i\alpha}\, , \label{Eqn_S3} \\
&\frac{1}{2\pi}\int_{0}^{2\pi}e^{\pm if_{1}(\phi)}d\phi \!=\!0, \label{Eqn_S4}  \\
&\frac{1}{2\pi}\int_{0}^{2\pi}e^{\pm if_{2}(\phi)}d\phi \!=\! 0, \label{Eqn_S5}  \\
&\frac{1}{2\pi}\int_{0}^{2\pi}e^{\pm i(f_1(\phi)-f_2(\phi))}d\phi \!=\!0.
\label{Eqn_S6}
\end{align}
Since, $\alpha$ is independent of $\phi$, we obtain $f_1(\phi)+f_2(\phi) \!=\!\alpha$ from from Eq.~\eqref{Eqn_S3}. From Eq.~\eqref{Eqn_S4}, Eq.\eqref{Eqn_S5} and Eq.\eqref{Eqn_S6}, we find that $f_{1,2}(\phi)$ should be linear functions of $\phi$, i.e., $f_{1(2)}(\phi)\!=\!\alpha_{1 (2)} + n_{1 (2)}\phi$, along with $\alpha_1 + \alpha_2 \!=\! \alpha$ (mod $2\pi$) and $n_1+n_2 \!=\! 0$,
where $n_1$ and $n_2$ are non-zero integers.

Without loss of generality, we use $\{\alpha_1, n_1\} \!=\!\{0, 1\}$ and $\{\alpha_2, n_2\} \!=\! \{\alpha, -1\}$, such that $f_1(\phi) \!=\! \phi$ and $f_2(\phi) \!=\! -\phi+\alpha$, and obtain the following expression.
\begin{align}
&G(q,r;m,m') \!=\! \avg{I_q(z,\phi)I_r(z,-\phi+\alpha)}_{\phi}  \nonumber\\
& =\frac{1}{4} \Big[ |U_{qm}(z)U_{rm}(z) +e^{i\alpha}U_{qm'}(z)U_{rm'}(z)|^2 \notag \\ &
+ I^{q}_m(z)I^{r}_{m'}(z) 
+ I^{r}_m(z)I^{q}_{m'}(z) \Big], 
\label{Eqn_S7}
\end{align}
where $I^{q}_m(z)$ is the normalized light intensity at waveguide $q$ after a propagation distance $z$ when light is launched only in the waveguide $m$ at $z \!=\! 0$.
Comparing Eq.~\eqref{Eqn_S1} and Eq.~\eqref{Eqn_S7}, we can write %
\begin{align}
\Gamma(q,r;m,m')& \!=\! 4G(q,r;m,m') \notag \\& 
- I^{q}_m(z)I^{r}_{m'}(z) - I^{r}_m(z)I^{q}_{m'}(z). 
\label{Eqn_S8}
\end{align}
In summary, the quantum mechanical observable, photon number correlation in Eq.~\eqref{Eqn_S1}, can be constructed from the generalized intensity correlation and intensity measurements. For such experiments, it is crucial to evolve {\it specific} initial states of laser light with a tunable relative phase.


\subsection{N-photon NOON states} \label{secS_N-photon_NOON_states}

In this section, we consider the evolution of the $N$-photon NOON state in a photonic lattice and then generalize the results of the previous section~\ref{secS_Two_photon_NOON_states}.
The NOON state, 
${\ket{\psi_{mm'}(N,\alpha)} \!=\! \dfrac{1}{\sqrt{2\cdot N!}}(\hat{a}^{\dagger N}_{m} +e^{-i\alpha}\hat{a}^{\dagger N}_{m'})\ket{0}}$, is initially coupled to the $m$-th and $m'$-th sites of the photonic lattice.
Here, we consider $N\leq M$, where $M$ is the total number of waveguides in the lattice.
The $N$ photons can come out from any $N$ lattice sites, represented by $\vec{s} \!=\! [s_1,s_2,...,s_N]$. Photon number correlation for the $N$-photon NOON states is defined here as
\begin{align}
&\Gamma(\vec{s}; \, m,m')  \notag \\&
=\frac{2}{N!}\bra{\psi_{mm'}(N,\alpha)}\creop{s_1}(z)\creop{s_2}(z)\cdots\creop{s_N}(z) \aniop{s_N}(z)\aniop{s_{N-1}}(z) \notag \\&
\cdots \aniop{s_1}(z)\ket{\psi_{mm'}(N,\alpha)} \notag\\
&=|U_{s_1m}(z)U_{s_2m}(z)\cdots U_{s_Nm}(z) \notag \\& 
+e^{i\alpha}U_{s_1m'}(z)U_{s_2m'}(z) \cdots U_{s_Nm'}(z)|^2, 
\label{Eqn_S9}
\end{align}
considering the commutation algebra for photons.

To obtain the photon number correlation in Eq.~\eqref{Eqn_S9} from intensity correlation measurements, we define an $N$-point generalized intensity correlation as below
\begin{eqnarray}
&G(\vec{s};m,m')(z) \!=\! \avg{\prod_{j=1}^{N}I_{s_j}(z,f_{j}(\phi))}_{\phi\in[0,2\pi]}  \, , \label{Eqn_S10}
\end{eqnarray}
where $I_{s_j}(z,f_{j}(\phi)) \!=\! \frac{1}{2}|U_{s_jm}+U_{s_jm'} e^{if_{j}(\phi)}|^2$ is the normalized intensity at site $s_j$ when light is launched at sites $m$ and $m'$ with relative phase $f_{j}(\phi)$. So  Eq.~\eqref{Eqn_S10} becomes 
\begin{align}
&G(\vec{s};m,m') 
= \frac{1}{2^N}\avg{\prod_{j=1}^{N} \Big( U^{*}_{s_j m}U_{s_j m} + U^{*}_{s_j m'}U_{s_j m'} \notag \\&
+ U^{*}_{s_j m}U_{s_j m'}e^{i f_{j}(\phi)} + U^{*}_{s_j m'}U_{s_j m}e^{-i f_{j}(\phi)} \Big)}_{\phi}\, . \, 
\label{Eqn_S11}
\end{align}
Notice that Eq.~\eqref{Eqn_S11} is a generalization of Eq.~\eqref{Eqn_S2} for $N$ points.
Comparing Eq.~\eqref{Eqn_S9} with Eq.~\eqref{Eqn_S11}, and following the same steps discussed in the previous section, we obtain
\begin{align}
&\sum_{j=1}^{N}f_j(\phi) \!=\! \alpha\, ,\,  \label{Eqn_S12} \\
& f_j(\phi) \!=\! \alpha_j + n_j\phi \, , {\text{where}} \quad j\in[1, N]\,  \label{Eqn_S13}
\end{align}
such that $\sum_{j=1}^{N}\alpha_j \!=\! \alpha$ (mod $2\pi$), $n_1\pm \sum_{j=2}^{l<N} n_j \ne 0$, and $\sum_{j=1}^{N}n_j\!=\!0$, where each $l\,(\in [2,\,N-1])$ value gives us one constraint.

Using Eqs.~\eqref{Eqn_S9} through \eqref{Eqn_S13}, and writing $\vec{s}$ as $\{p, q, r\}$ for $N\!=\!3$, we obtain
\begin{align}
&G(p,q,r;m,m') \!=\! \frac{1}{8}[|U_{pm}U_{qm}U_{rm}+e^{i\alpha}U_{pm'}U_{qm'}U_{rm'}|^2 \notag \\
&+ |U_{pm'}U_{qm}U_{rm}|^2+|U_{pm}U_{qm'}U_{rm'}|^2+|U_{pm}U_{qm'}U_{rm}|^2 \notag\\
&+|U_{pm'}U_{qm}U_{rm'}|^2+|U_{pm}U_{qm}U_{rm'}|^2+|U_{pm'}U_{qm'}U_{rm}|^2]\notag\\
&=\frac{1}{8}[\Gamma(p,q,r;m,m')+I_{m'}^pI_m^qI_m^r+I_m^pI_{m'}^qI_{m'}^r+I_m^pI_{m'}^qI_m^r \notag \\
&+ I_{m'}^pI_m^qI_{m'}^r+I_m^pI_m^qI_{m'}^r+I_{m'}^pI_{m'}^qI_m^r],  
\label{Eqn_S14}
\end{align}
where we have used 
$\{\alpha_1, \, n_1 \} \!=\! \{0, 1\}$,
$\{\alpha_2, \, n_2 \} \!=\! \{0, 2\}$, and 
$\{\alpha_3, \, n_3 \} \!=\! \{\alpha, -3\}$,
such that $f_1(\phi) \!=\! \phi$, $f_2(\phi) \!=\! 2\phi$ and $f_2(\phi) \!=\! -3\phi+\alpha$, without any loss of generality. Eq.~\eqref{Eqn_S14} can be rearranged to obtain Eq.~\eqref{Eqn_4} 
~describing the constructed quantum correlation for three-photon NOON states. It should be noted that Eq.~\eqref{Eqn_S11} contains $2^N -1$ terms on the right-hand side. This can also be noticed for specific cases of $N\!=\!2$ and $3$ in Eq.~\eqref{Eqn_S8} and Eq.~\eqref{Eqn_S14}, respectively.


\subsection{Two-particle anyonic correlation} \label{secS_Anyonic correlation}
For Abelian anyons, the generalized commutation relations~\cite{keilmann2011statistically}, in one-dimension, are
\begin{align}
\aniop{j} \creop{k} - e^{-i\alpha\,\operatorname{sgn}(j-k)} \creop{k} \aniop{j} &= \delta_{jk}, \notag\\
\aniop{j} \aniop{k} - e^{i\alpha\,\operatorname{sgn}(j-k)} \aniop{k} \aniop{j} &= 0, \notag \\
\creop{j} \creop{k} - e^{i\alpha\,\operatorname{sgn}(j-k)} \creop{k} \creop{j} &= 0,
\label{Eqn_S15_n}
\end{align}
where \(\alpha\) is the anyonic exchange phase and $\operatorname{sgn}()$ is the sign function. Using these relations, 
the only non-vanishing contributions to the two-particle amplitude arise from the direct ($q \rightarrow m$, $r \rightarrow m'$) and exchange processes ($q \rightarrow m'$, $r \rightarrow m$), yielding
\begin{align}
\aniop{r} \aniop{q} \creop{m} \creop{m'} |0\rangle
& \!=\!
\left[
\delta_{qm}\delta_{rm'}
\!+\!
e^{i\alpha\,\operatorname{sgn}(m-m')}\delta_{qm'}\delta_{rm}
\right]|0\rangle.
\label{Eqn_S16_n}
\end{align}
Accordingly, the quantum correlation of two indistinguishable anyons reduces to
\begin{align}
&\Gamma^{A}(q,r;m,m')\!=\! \nonumber \\ 
& \left|
U_{qm}(z)U_{rm'}(z)
\!+\!
e^{i\alpha\,\operatorname{sgn}(m-m')}U_{qm'}(z)U_{rm}(z)
\right|^2.
\label{Eqn_S17_n}
\end{align}
Equation~\eqref{Eqn_S17_n} shows that the anyonic statistics enter only through the relative phase between the direct and exchange paths.

Similar to Sec.~\ref{secS_Two_photon_NOON_states}, we define the output intensities $I_q(f_1,z)$, $I_r(f_2,z)$ 
and their phase-averaged correlation $G^A(q,r;m,m') \!=\!
\avg{I_q(z,f_1)I_r(z,f_2)}_{\phi}$. 
To emulate $\Gamma^{A}$ using intensity correlations, we compare Eq.~\eqref{Eqn_S17_n} with the expression of $G^A(q,r;m,m')$, and constrain the phases as
\begin{align}
&f_1(\phi)-f_2(\phi)=\alpha \operatorname{sgn}(m-m'). \qquad \label{Eqn_S18_n}
\end{align}
So the anyonic intensity correlation is given by
\begin{align}
&G^A(q,r;m,m') -
I_m^q(z)I_{m}^r(z)
-
I_{m'}^r(z)I_{m'}^q(z) \notag \\
&=\frac{1}{4}\Big[
|U_{qm}(z)U_{rm'}(z)+e^{i\alpha \operatorname{sgn}(m-m')}U_{qm'}(z)U_{rm}(z)|^2 \Big] \notag\\
\label{Eqn_S19_n}
\end{align}
In summary, the quantum correlations of two anyons in one dimension can be experimentally emulated using Eq.~\eqref{Eqn_S19_n} along with the phase constraint in Eq.~\eqref{Eqn_S18_n}.

%

\begin{figure}[b!]
    \centering
\includegraphics[width=0.9\linewidth]{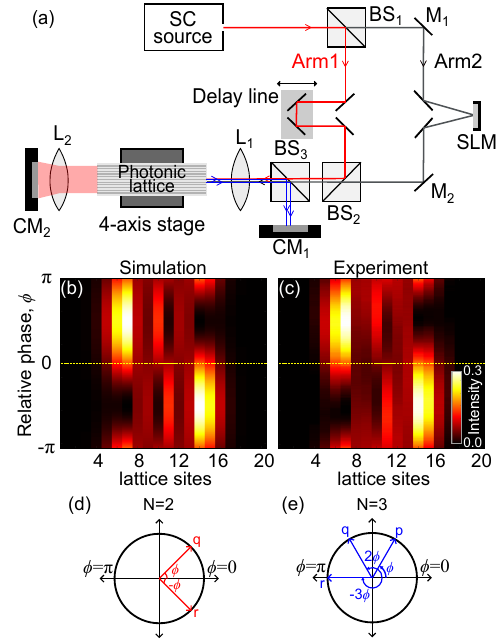}
    \caption{(a) A schematic of the experimental setup.
    (b) Numerically calculated intensity profiles distributed among the $M \!=\!20$ lattice sites of a one-dimensional lattice as the relative phase is varied from $\pi$ to $-\pi$. 
    (c) Experimentally measured output intensity as a function of the input phase, controlled via the SLM. The reference condition $\phi \!=\! 0$ is indicated, relative to which all other phase values are calibrated.
    (d, e) Schematics showing how the phase values are chosen to construct the intensity correlation for $N \!=\!2$ and $N\!=\!3$ NOON states, respectively.
    }
\phantomsection
\label{Fig_S_experimental_details}
\end{figure}


\section{Fabrication details}
\label{secS_Fabrication details}
We fabricate waveguides and waveguide arrays in borosilicate (BK7) glass using femtosecond laser writing~\cite{davis1996writing, shit2024probing}. These waveguide structures are created at a depth of $100 - 150\, \mu$m from the top surface of the glass using $260$~fs laser pulses at $500$~kHz repetition rate, generated from a
commercially available Yb-doped fiber laser system (Satsuma, Amplitude Laser Inc.). The fabrication process was optimized to realize low-loss single-mode waveguides operating near the wavelength range of $930$~nm to $1064$~nm. The propagation loss for this wavelength range was estimated to be $0.31$~dB/cm to $0.41$~dB/cm. We perform all characterizations using horizontally polarized light, generated from a wavelength-tunable super-continuum
source (NKT Photonics). The fundamental modes supported by the waveguides are elliptical in shape. The measured mode field diameters ($1/e^2$ of the intensity peak) along the vertical and horizontal axes are $21.5 \, \mu$m and $20.9 \, \mu$m, respectively, at $930$~nm.

\section{Details on intensity correlation measurement}
\label{secS_Details on intensity correlation measurement}

\subsection{Measurement protocol}
\label{Measurement_protocol}
In our experiments, the preparation of specific initial states, their evolution through a photonic lattice, and the measurement of intensity profiles are carried out using the following steps.

(a) We generate wavelength-tunable coherent states of light using a super-continuum source (SCS) along with an acousto-optic tunable filter (NKT Photonics).  
The light beam from the SCS is split into two arms using a $50\!:\!50$ beam splitter (BS$_1$), as shown in Fig.~\ref{Fig_S_experimental_details}(a). 
In arm 1, the beam passes through a variable delay line, whereas in arm 2, it is reflected by a spatial light modulator (SLM), which controls the relative phase $\phi$ of the light in the two arms.  
The beams are then recombined at a second $50\!:\!50$ beam splitter (BS$_2$) and focused onto the desired lattice sites ($m$ and $m'$) using a bi-convex lens (L$_1$).
The glass wafer containing the photonic lattices is mounted on a
$4$-axis translation stage for precise light coupling. Back-reflected light from the input facet of the glass wafer is imaged on a camera (CM$_1$) using a beam splitter (BS$_3$). The intensity profile at the output of the lattice is imaged on a CMOS camera (CM$_2$) using a bi-convex lens (L$_2$).

(b) The output intensity distribution of the photonic lattice is sensitive to the initial relative phase $\phi$. 
To calibrate the SLM and to find out its configuration corresponding to $\phi \!=\! 0$, we measure intensity distributions at the output of the one-dimensional array as a function of the voltage applied to the SLM pixels. Comparing the experimentally and numerically obtained intensity distributions, we can identify the SLM voltage configuration corresponding to $\phi \!=\! 0$, see Fig.~\ref{Fig_S_experimental_details}(b,c).

(c) We measure output intensity patterns across all waveguides for $256$ values of $\phi$ that are uniformly spaced from $0$ to $2\pi$. 
Then the intensity correlation $G$ for a specific $\alpha$-value can be obtained by phase-averaging the product of the intensities at different lattice sites -- the integration in Eq.~\eqref{Eqn_2}
(main text) is replaced with a summation over $256$ phase-points.
For example, in the case of $N \!=\!2$ NOON state, we perform phase averaging of $I_q(\phi) I_r(-\phi+\alpha)$ to obtain $G$. 
Figures~\ref{Fig_S_experimental_details}(d, e) illustrate the selection of phases for two-point and three-point correlation cases, respectively.
The quantum correlation matrix $\Gamma$ is then constructed using the intensity correlation, as discussed in Sections~\ref{secS_Two_photon_NOON_states}, \ref{secS_N-photon_NOON_states}.

As mentioned before, we obtain intensity correlations, as defined in Eq.~\eqref{Eqn_2} for $N\!=\!2$, by replacing the integral with a summation over $n_{\phi}$ number of equally spaced phase points. Our numerics suggest that the phase resolution should be $n_{\phi} \geq 2^N -1$ for emulating the $N$-photon correlation (Appendix~\ref{Robustness}). However, in the presence of phase fluctuations, a higher phase resolution would increase the accuracy of the protocol. For all our experiments, we have used a phase resolution of $n_{\phi}\!=\!256$ that is determined by the SLM.


\begin{figure}[t]
    \centering
\includegraphics[width=1.0\linewidth]{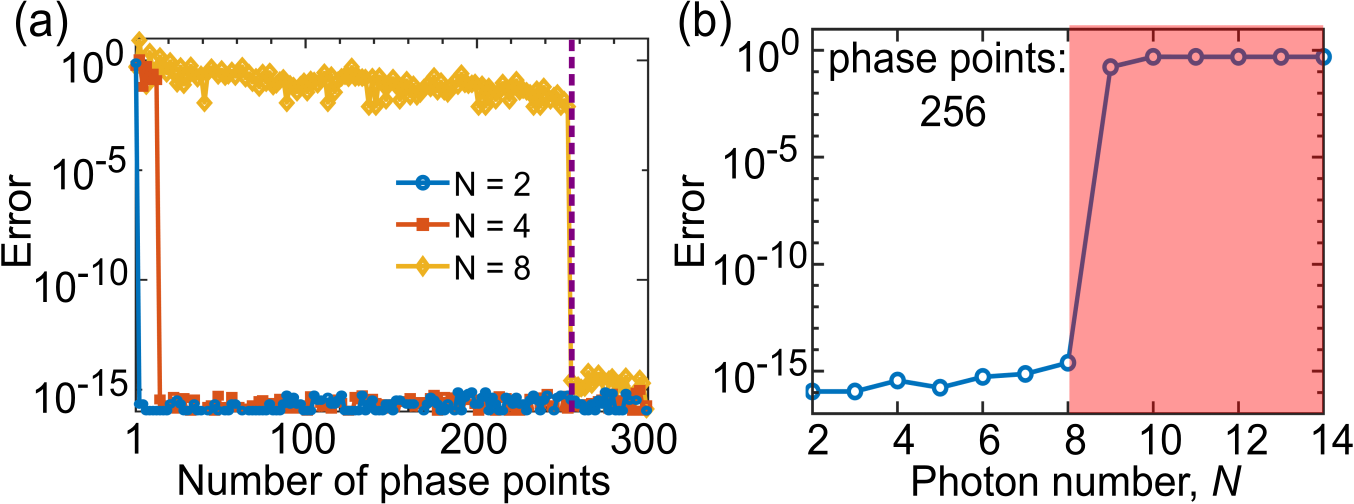}
    \caption{(a) Numerically obtained error as a function of the number of phase points. The phase resolution in our experiment is indicated by the dashed line. (b)  Error as a function of $N$ with a fixed number of phase points, i.e., 256. The error increases sharply beyond $N\!=\!8$, due to an insufficient number of phase points.
    }
\phantomsection
\label{Fig_S_Robustness}
\end{figure}

\subsection{Wavelength tuning}
\label{Wavelength_tuning}
Light evolution in the straight photonic lattice is governed by the normalized propagation distance $Jz$. 
In our experiments, the maximal propagation distance of the photonic lattice
is fixed ($z \!=\! 40$~mm), and we only have access to the output intensity profiles. In this situation, we vary the wavelength of light $\lambda$ to tune the coupling $J$, and hence, the normalized propagation distance~\cite{sinha2025probing}. The coupling $J$ varies almost linearly in the wavelength range of interest ($930$~nm to $1064$~nm), and this wavelength-tuning protocol allows us to observe the dynamics of light as a function of $Jz$, see Fig.~\ref{Fig_1}
in the main text.

\begin{figure}[t]
    \centering
\includegraphics[width=1.0\linewidth]{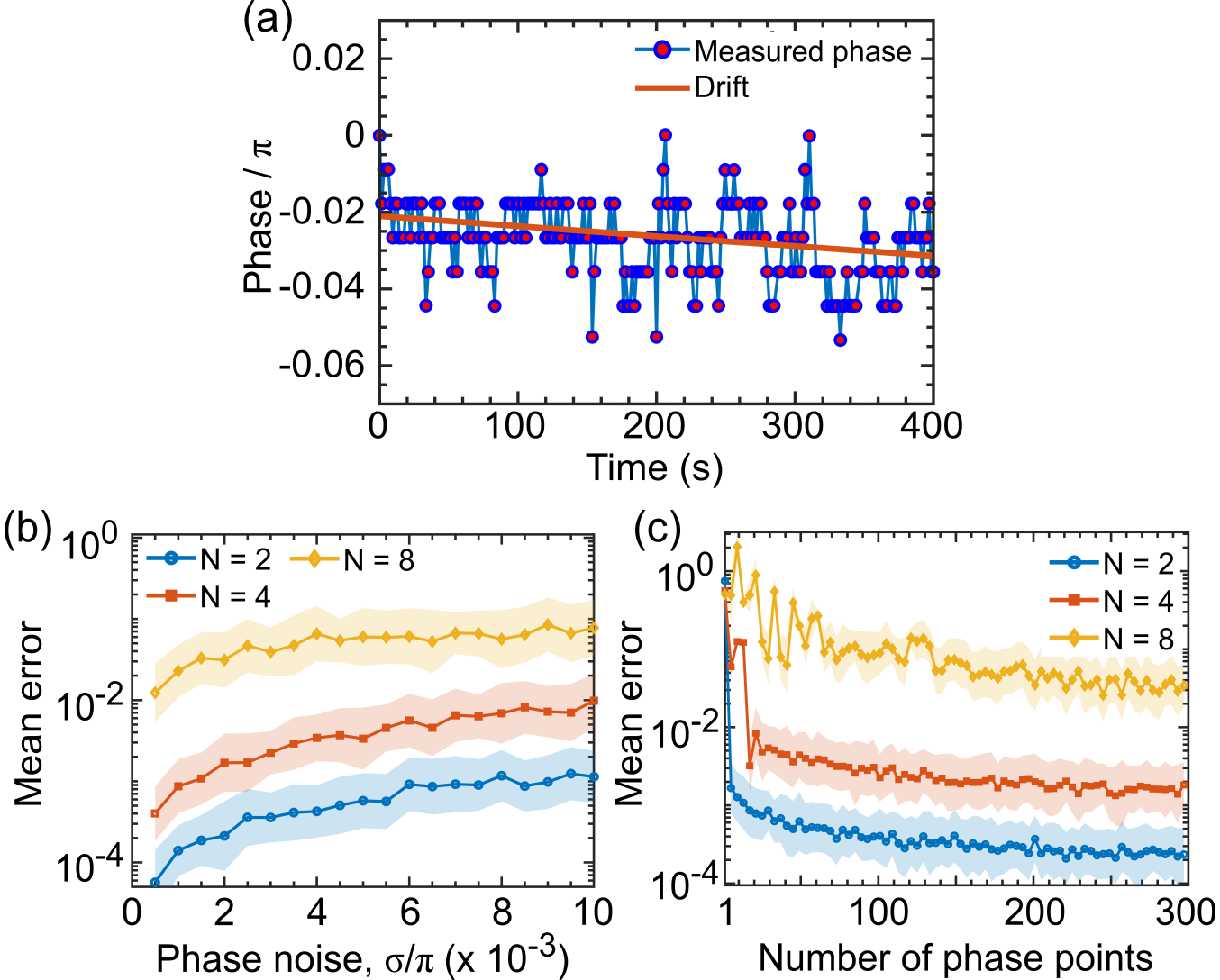}
    \caption{(a) Experimental phase fluctuations measured over $400~\mathrm{s}$, showing a total drift of $0.01\pi$. 
    The standard deviation of the phase fluctuation is $0.005\pi$.
    (b) Mean error versus phase noise, calculated for $256$ phase points . The shade indicates the standard deviation.
    (c) Mean error versus phase points for a typical phase fluctuation with $\sigma \!=\! 0.002\, \pi$. For (b, c), we have considered $50$ noise realizations.
    }
\phantomsection
\label{Fig_S_Error}
\end{figure}

\subsection{Influence of the resolution and noise in phase} \label{Robustness}

In this section, we first find out the phase resolution (i.e., the number of phase points $n_{\phi}$ in the range of $[0, \, 2\pi]$) required to accurately emulate $N$-photon quantum correlations. To this end, we consider the flat-band rhombic lattice and define error as
$\text{Err}=|P_L^{I}-P_L^{Q}|$, where $P_L^{I}$ and $P_L^{Q}$ denote the localization probabilities obtained from intensity correlations and quantum correlations, respectively. 
As shown in Fig.~\ref{Fig_S_Robustness} (a), for $N\!=\!2,4$, and $8$, the error goes to zero when the number of phase points is more than $2^N -1$. In other words, with the experimentally achieved 
phase points of $256 \, (= 2^{8}$),
the error remains negligible up to $N\!=\!8$; however, it sharply increases beyond this point; see Fig.~\ref{Fig_S_Robustness}(b). The behavior in Fig.~\ref{Fig_S_Robustness}(b) is consistent with the trend in Fig.~\ref{Fig_S_Robustness}(a). The discussion in Sec.~\ref{secS_N-photon_NOON_states} [Eq.~\eqref{Eqn_S12}, Eq.~\eqref{Eqn_S13}] and the associated constraints also indicate that the emulation of an $N$-photon NOON state requires at least $2^N -1$ uniformly spaced phase points in the range of $[0, \, 2\pi]$. For fewer phase points, the protocol can not accurately reproduce the corresponding quantum correlations.

In the experiment, various environmental factors can cause small fluctuations and a drift in the relative phase of the input state. We characterize the phase stability by monitoring the interference fringes generated by the light waves from the two arms, as shown in Fig.~\ref{Fig_S_experimental_details}(a). A total phase drift of $0.01\pi$ was observed over $400~\mathrm{s}$ [Fig.~\ref{Fig_S_Error}(a)]. Note that each intensity correlation experiment requires approximately 
$70~\mathrm{s}$, which corresponds to a small mean drift of $0.002\pi$. Phase noise was quantified by fitting its distribution with a Gaussian, yielding a standard deviation of $0.005\pi$. This standard deviation lies within the experimental phase resolution of $\pi/128$. 

To investigate how the accuracy of our protocol is influenced by the phase noise, we introduce random noise in the input relative phases in Eq.~\eqref{Eqn_2}. We then calculate the mean and standard deviation of the error $\text{Err}$ considering $50$ noise realizations.
Fig.~\ref{Fig_S_Error}(b) shows the mean error as a function of the phase noise for a fixed $256$ phase points. 
Whereas, Fig.~\ref{Fig_S_Error}(c) shows the mean error as a function of phase points, for a fixed phase fluctuation of $\sigma\!=\!0.002\, \pi$. 
Note that the mean error increases with the phase noise and decreases as the number of phase points increases.

The above results show that, in principle, the protocol can be extended to a large $N$; however, in practice, it is limited by phase resolution and stability of the SLM as well as by the experimental noise.

\begin{figure*}[hbt!]
    \centering
\includegraphics[width=0.9\linewidth]{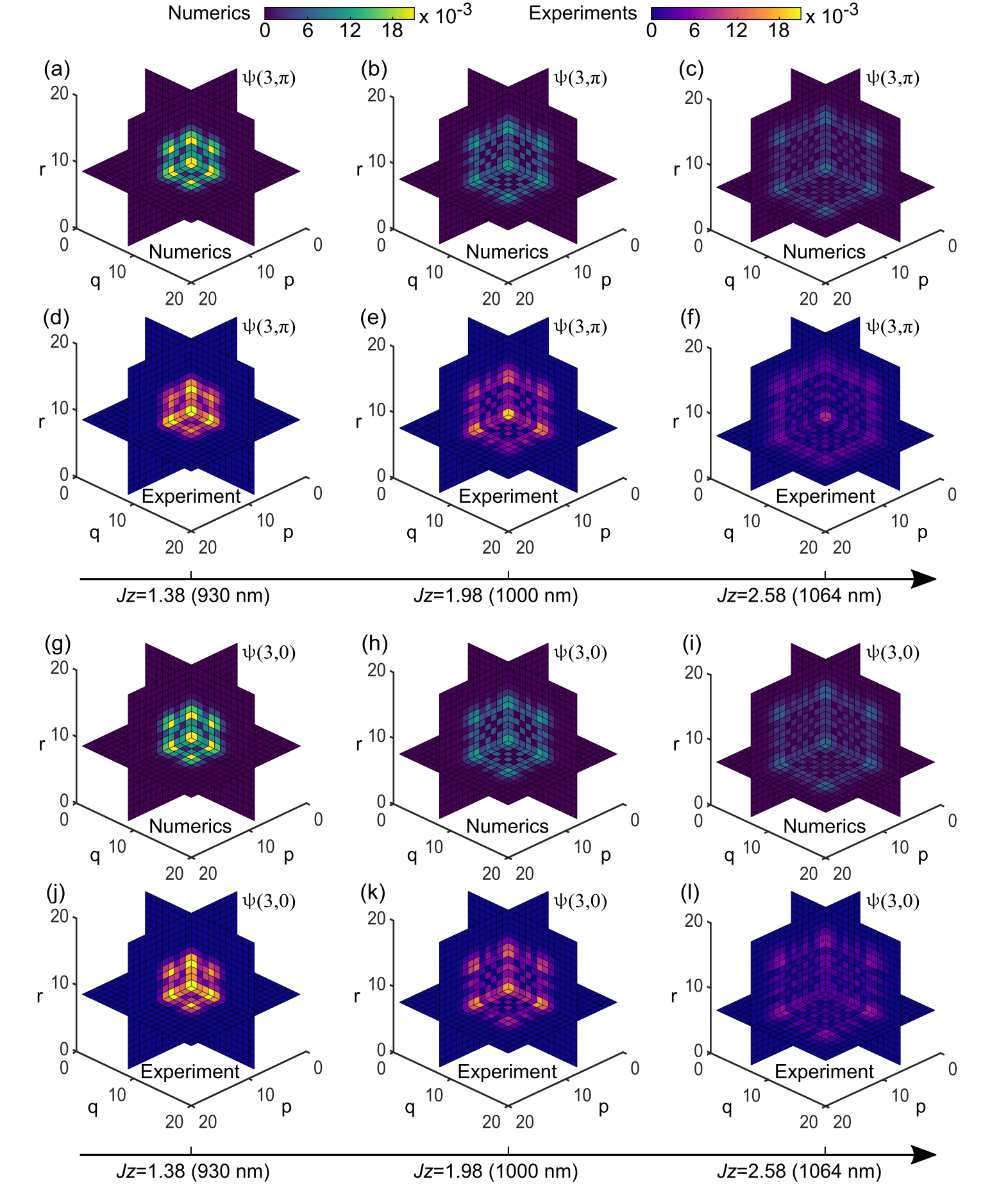}
    \caption{(a-c) Photon number correlations ($\Gamma$) for three-photon NOON state $\psi_{10,11}(3, \alpha\!=\!\pi)$ at three different effective propagation distances $Jz$. (d-f) Experimentally constructed $\Gamma_{\text{expt}}$ associated with (a-c). (g-l) Same as (a-f), for $\psi_{10,11}(3, \alpha\!=\!0)$. Notice that both $\alpha\!=\!\pi$ and $0$ produces the same $\Gamma$ in this case. The coordinate
planes for the three columns are $\{9, 9, 9\}$, $\{8, 8, 8\}$ and $\{7, 7, 7\}$, respectively.
    }
    \phantomsection
\label{Fig_1D_3_photon}
\end{figure*}

\begin{figure*}[]
\centering
\includegraphics[width=1\linewidth]{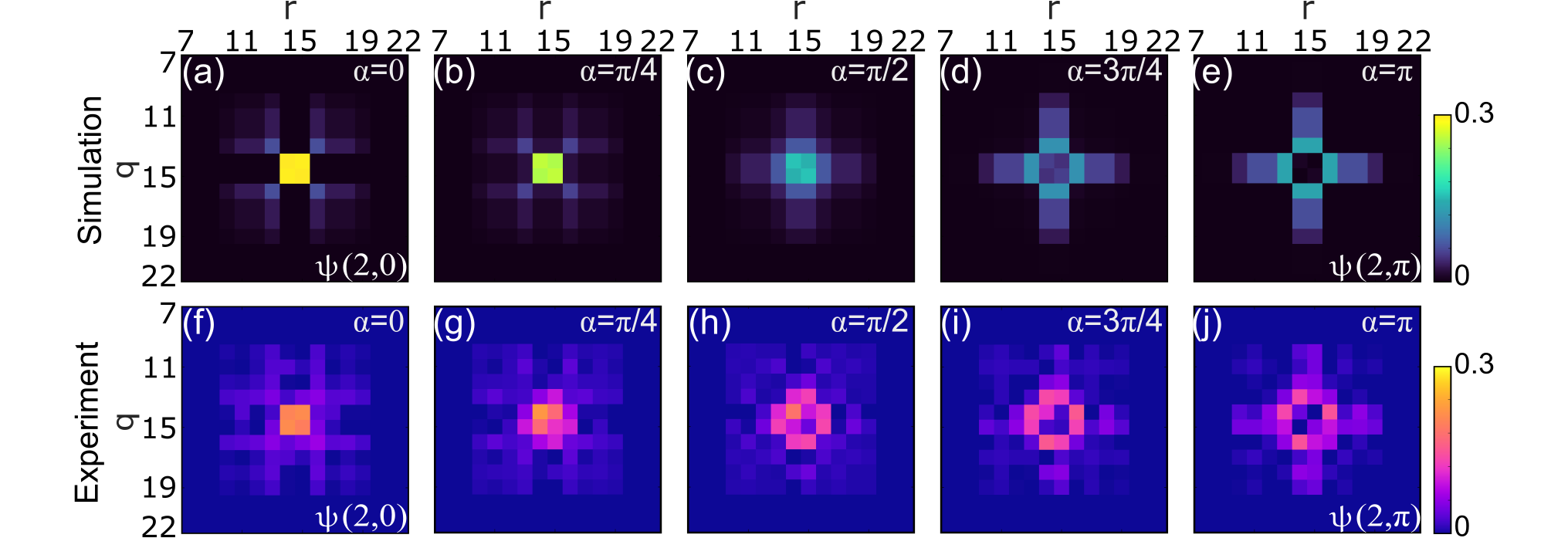} 
\caption{Experimental results showing the capability of tuning the NOON state phase $\alpha$. {(a-e) Numerically obtained photon number correlation at $Jz\!=\!0.91$ for two-photon NOON states with five different phase values $\alpha$ indicated in each image. (f-j) Experimental data corresponding to (a-e) obtained from intensity correlation measurements.}
}
\phantomsection
\label{Fig_S_varying_alpha}
\end{figure*}

\section{Three-photon NOON state correlations in the one-dimensional lattice}\label{secS_with_1D_3_photon}

In the main text, we have presented bunching and anti-bunching of two-photon NOON states. In this section, we consider three-photon NOON states evolving in the one-dimensional lattice shown in Fig.~\ref{Fig_1}(c, d). Figures~\ref{Fig_1D_3_photon}(a-c) present numerically calculated
photon number correlations $\Gamma(p,q,r; 10, 11)$ for the initial state $\psi_{10,11}(3, \alpha\!=\!\pi)$. 
Here, we consider three different effective propagation distances $Jz$ that were realized experimentally. Note that the coordinate
planes for (a-c) are $\{9, 9, 9\}$, $\{8, 8, 8\}$ and $\{7, 7, 7\}$, respectively.
Using the intensity correlation protocol Eq.~\eqref{Eqn_4}, we 
then constructed $\Gamma_{\text{expt}}$, shown in  Figs.~\ref{Fig_1D_3_photon}(d-f), that are in good agreement with Figs.~\ref{Fig_1D_3_photon}(a-c).

For the three-photon NOON state with zero phase, i.e., $\psi_{10,11}(3, \alpha\!=\!0)$, we calculated and experimentally constructed the correlations, as shown in Figs.~\ref{Fig_1D_3_photon}(g-l). It is worth mentioning that both $\alpha \!=\! \pi$ and $\alpha \!=\! 0$ produce the same $\Gamma$ in the case of $N\!=\!3$ NOON state evolving in a one-dimensional lattice. To further explain this behavior,  
we consider 
the Bloch modes (in momentum-space) with maximum group velocity around $ka\!=\!\pm\pi/2$, as discussed in the main text.
In this case, the states with $\alpha \!=\! \pi$ and $0$ can be expressed in momentum space as 
\begin{align}
&\psi(3,\alpha \!=\! 0/\pi) \!=\! 
\frac{1}{4\sqrt{3!}}\Big[ \hat{a}_{\pi/2}^{\dagger 3} \, e^{-i\frac{3\pi}{2}m} (1\mp i) \notag \\
&+ \hat{a}_{-\pi/2}^{\dagger 3} \, e^{i\frac{3\pi}{2}m} (1\pm i) + 
3\, \hat{a}_{\pi/2}^{\dagger 2}\hat{a}_{-\pi/2}^{\dagger} \, e^{-i\frac{\pi}{2}m} (1\mp i) \notag \notag \\
&+ 3\, \hat{a}_{\pi/2}^{\dagger}\hat{a}_{-\pi/2}^{\dagger 2} \, e^{i\frac{\pi}{2}m} (1\pm i) \Big], 
\label{Eqn_S20}
\end{align}
where the upper (lower) signs corresponds to $\alpha\!=\!0$ ($\pi$). 
Note that the modulus square of the coefficients for the four terms in Eq.~\eqref{Eqn_S20} are equal irrespective of $\alpha\!=\!0$ or $\pi$. This qualitatively explains the prominent corner lobes in Figs.~\ref{Fig_1D_3_photon}.

\section{Two-photon NOON state correlations in the rhombic lattice with varying NOON phase $\alpha$.}
\label{secS_with_varying_NOON_phase}

To experimentally simulate the two-photon NOON state $\psi_{14,15}(2,\alpha)$ with a variable phase $\alpha$, we follow the protocol described in Appendix~\ref{secS_Details on intensity correlation measurement}.
However, $f_1$ and $f_2$ [see Eq.~\eqref{Eqn_2}
] are now selected from phase points corresponding to $\phi$ and $-\phi \!+\! \alpha$, respectively.
The resulting intensity correlation enables us to emulate the output photon number correlation $\Gamma$ of NOON states with a phase $\alpha$.
Fig.~\ref{Fig_S_varying_alpha} presents experimentally obtained $\Gamma$ for the two-photon NOON state in the flat-band rhombic lattice. Here, the variation of $\alpha$ from $0$ to $\pi$ alters the localization feature in the correlation matrix to delocalization. This procedure naturally generalizes to high-NOON states by selecting intensity profiles at appropriately chosen phase points, such that their phase-averaged product yields the desired joint intensity correlation.  
This intensity correlation can then be mapped to the corresponding photon-number correlation using Eq.~\eqref{Eqn_S8}.

\begin{figure}[t]
\centering
\includegraphics[width=0.85\linewidth]{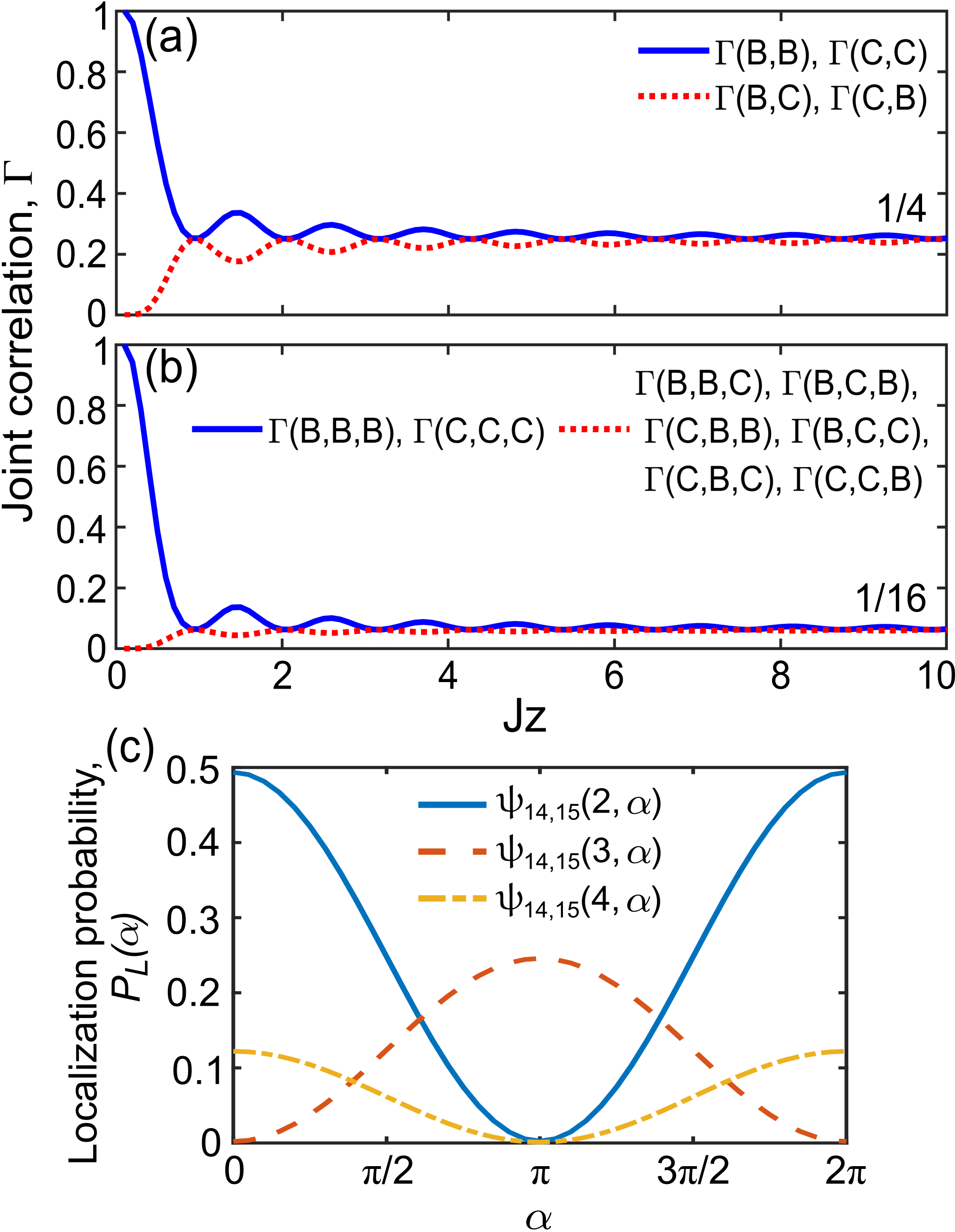} 
\caption{(a) Variation of joint correlation for the two-photon state $\psi_{14,15}(2,0)$ launched at the $B$ and $C$ sites. After some finite propagation, $\Gamma$ values approach $1/4$.
(b) Same as (a) for the three-photon state $\psi_{14,15}(3,\pi)$. In this case, $\Gamma$ values approach $1/16$.
Here, $\Gamma(B, C)$ denotes the joint correlation of photons detected 
at the output B site ($q\!=\!14$) and C site ($r\!=\!15$); similar notation is used for other site combinations.
(c) %
Variation of localization probability $P_L$ as a function of the phase $\alpha$ for $N\!=\!2, \, 3$ and $4$.
}
\phantomsection
\label{Fig_S_localization_oscillation}
\end{figure}

\section{NOON state in the flat-band rhombic lattice}
\label{secS_analysis of NOON state in flat-band}
In the main text, we demonstrated that the localization-delocalization of multi-photon NOON states in a flat band rhombic lattice can crucially depend on the phase $\alpha$ of the NOON states as well as the photon number $N$. In this section, we provide a detailed explanation of such behavior.

To obtain the localization probability $P_L$ of all NOON state photons occupying the flat band, we numerically evolve $\Gamma$ considering a large system size. 
Figure~\ref{Fig_S_localization_oscillation}(a) shows the $z$-evolution of $\Gamma_{q, r}$ and $q,r\in\{m,m'\}$ for $\psi_{m,m'}(2, 0)$. 
After some initial oscillations, all four correlation elements saturate to $1/4$.  
In the limit of long propagation distances, we can write the probability for $N\!=\!2$ as $P_L\!=\!\frac{1}{2}\! \sum \int {\text{d}z}\, \Gamma_{q, r}(z)$, where the summation runs over $q,r\in\{m,m'\}$ and the integration in performed to obtain a $z$-averaged value.
Similarly, for the $\psi_{m,m'}(3, \pi)$ case, all eight correlation elements saturates to $1/16$, resulting in $P_L\!=\!1/4$, see Fig.~\ref{Fig_S_localization_oscillation}(b). 
Figure~\ref{Fig_3}
in the main text presents $P_L$ up to $N\!=\!8$, alternately considering $\alpha\!=\!0$ and $\pi$ -- this clearly shows its dependence on $N$ as $2^{1-N}$. In experiments, the elements of the correlation matrix are obtained %
at $Jz\!=\!0.91$. Due to this finite propagation, the error in experimentally estimating $P_L$ is less than $2\, \%$.

\begin{figure}[]
\centering
\includegraphics[width=1\linewidth]{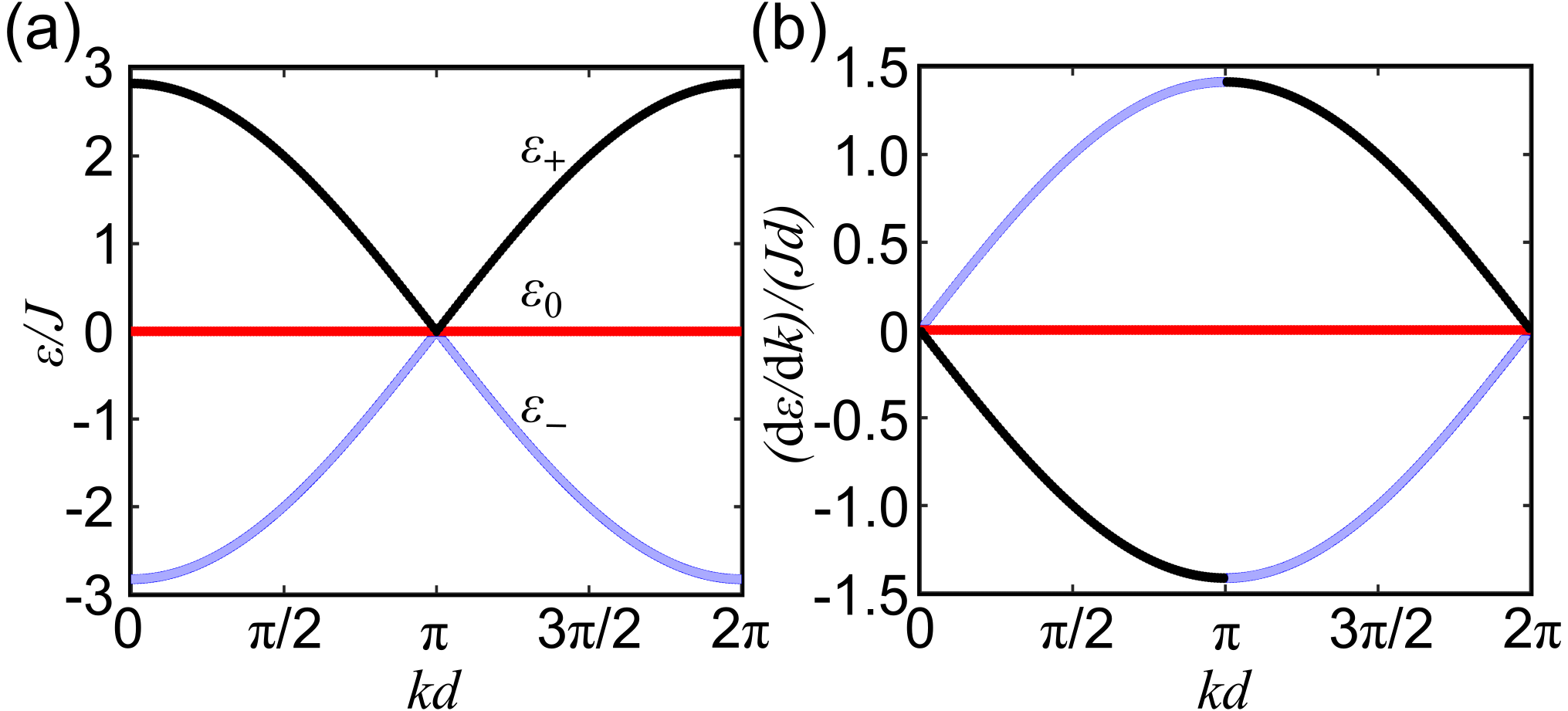} 
\caption{{(a) Band structure of the rhombic lattice. Notice that the middle band is perfectly flat at all $k$ values. (b) Group velocity as a function of $kd$.}
}
\phantomsection
\label{Fig_S_Spectrum}
\end{figure}

We now provide an explanation of the localization-delocalization of the correlation by expressing the initial states in the $k$-space basis. As discussed before, the spectrum of the rhombic lattice consists of a flat band $\varepsilon_{0}(k)\!=\!0$ and two dispersive bands $\varepsilon_{\pm}(k)\!=\!\pm 2J\sqrt{1+\cos(kd)}$, where $d$ is the lattice constant; 
see Fig.~\ref{Fig_S_Spectrum}(a). 
The eigenmodes of the flat and dispersive band(s) are given by $~(0,~1,~-1)^T/\sqrt{2}$ and $(\pm g(k),~1,~1)^T/\sqrt{2+g^2(k)}$, where $g(k)\!=\! \big(2(1+e^{-ikd})/(1+e^{ikd})\big)^{1/2}$, respectively. Also note that the flat-band eigenmodes do not depend on $k$.
As shown in Fig.~\ref{Fig_S_Spectrum}(b) the group velocity $v\!=\!{\text{d}} \varepsilon/{\text{d}}k$ of the dispersive modes is maximal near $kd \!=\! \pi$; hence, to obtain an intuitive picture of the localization-delocalization phenomenon, we first perform the analysis at $kd \!=\! \pi+\delta$, where $\delta$ is a small positive number.
By denoting the creation operators of the flat band and the two dispersive bands (near $kd\!=\!\pi$) by $\hat{L}^{\dagger}$ and $\hat{D}^{\dagger}_{\pm}$, respectively, 
the real-space creation operators at the B and C site can be expressed as $\hat{a}^{\dagger}_B \!=\! (\sqrt{2}\hat{L}^{\dagger}+\hat{D}^{\dagger}_++\hat{D}^{\dagger}_-)/2$ and $\hat{a}^{\dagger}_C \!=\! (-\sqrt{2}\hat{L}^{\dagger}+\hat{D}^{\dagger}_++\hat{D}^{\dagger}_-)/2$, respectively. The two-photon NOON states considered in Fig.~2~%
can then be written as 

\begin{eqnarray}
    &\psi_{m,m'}(2,\alpha\!=\!0) \!=\! \frac{1}{2}(\hat{a}^{\dagger 2}_B + \hat{a}^{\dagger 2}_C)\ket{0} = \nonumber \\&
    \frac{1}{2\sqrt{2}} (\ket{2\hat{D}_+} + \ket{2\hat{D}_-}) + 
    \frac{1}{2}\ket{\hat{D}_+,\hat{D}_-} + \frac{1}{\sqrt{2}} \ket{2\hat{L}}, \label{Eqn_S21}
\end{eqnarray}
\begin{eqnarray}   
    &\psi_{m,m'}(2, \alpha\!=\!\pi) \!=\! \frac{1}{2}(\hat{a}^{\dagger 2}_B - \hat{a}^{\dagger 2}_C)\ket{0} \hfill{} \nonumber \\&
    =\frac{1}{\sqrt{2}}(\ket{\hat{D}_+,\hat{L}} +\ket{\hat{D}_-,\hat{L}}) . \label{Eqn_S22}
\end{eqnarray}
For the $\psi_{m, m'}(2,0)$ state in Eq.~\eqref{Eqn_S21}, the coefficients of the first two terms give the probability of both photons moving in either positive or negative direction, which is $1/8$. Similarly, the probability of one photon moving in the positive direction and the other one in the negative direction is $1/4$. Importantly, the last term in Eq.~\eqref{Eqn_S21} gives the probability of both photons in the flat band, which is $P_L\!=\!1/2$. 
On the other hand, for the $\psi_{m,m'}(2,\pi)$ in Eq.~\eqref{Eqn_S22}, the probability for both photons to be localized is zero. Evidently, Eqs.~\eqref{Eqn_S21} and \eqref{Eqn_S22} suggest that localization of both photons at the initial launching site is expected for phase $\alpha\!=\!0$, as observed in Figs.~2~
(g, h).

To explain the flipping of the above localization-delocalization for $N\!=\!3$ NOON states, we express these states as
\begin{eqnarray} 
    &\psi_{m,m'}(3,\alpha\!=\!0) \!=\! \frac{1}{2\sqrt{3}}(\hat{a}^{\dagger 3}_B + \hat{a}^{\dagger 3}_C)\ket{0} \nonumber \\&
    =\frac{1}{4\sqrt{2}}(\ket{3\hat{D}_+} +\ket{3\hat{D}_-})+ \frac{\sqrt{6}}{8}(\ket{2\hat{D}_+,\hat{D}_-}+ \nonumber \\&
    \ket{2\hat{D}_-,\hat{D}_+}) + \frac{\sqrt{6}}{4}(\ket{\hat{D}_+,2\hat{L}} +\ket{\hat{D}_-,2\hat{L}})\, ,\label{Eqn_S23}
\end{eqnarray}
\vspace{-2mm}
\begin{eqnarray} 
    &\psi_{m,m'}(3,\alpha\!=\!\pi) \!=\! \frac{1}{2\sqrt{3}} (\hat{a}^{\dagger 3}_B - \hat{a}^{\dagger 3}_C)\ket{0} \!=\! \frac{\sqrt{3}}{4} 
    (\ket{2\hat{D}_+,\hat{L}} \nonumber \\&
    + \ket{2\hat{D}_-,\hat{L}}) + \frac{\sqrt{6}}{4} \ket{\hat{D}_+,\hat{D}_-,\hat{L}} + \frac{1}{2}\ket{3\hat{L}}.
    \label{Eqn_S24}
\end{eqnarray}
%
Note that the probability for all three photons to be localized at the flat band is zero for $\psi_{m,m'}(3,0)$ in Eq.~\eqref{Eqn_S23}. However, this probability for $\psi_{m,m'}(3,\pi)$ in Eq.~\eqref{Eqn_S24} is $P_L\!=\!1/4$. The above analysis near $kd\!=\!\pi$ qualitatively explains the localization-delocalization features observed in Figs.~\ref{Fig_2}
(e-l). We note that it is straightforward to generalize the above analysis for the $N$ photon NOON state. 
This approximate calculation performed near $kd\!=\!\pi$ gives the exact values of $P_L$ due to the interesting fact that the coefficient of $\ket{N\hat{L}}$ does not depend on $k$. In this context, we note that the Bloch modes of dispersive bands at $kd \!=\!0$ also have zero group velocity; however, their contribution to the localization probability is insignificant in the thermodynamic limit.

To obtain the dependence of $P_L$ on the phase $\alpha$, the NOON state %
$\psi_{m,m'}(N,\alpha)$ can be expressed in terms of $\hat{L}^{\dagger}$ and $\hat{D}^{\dagger}_{\pm}$. As in Eq.~\eqref{Eqn_S21}-Eq.~\eqref{Eqn_S24}, we obtain the coefficient of $\ket{N\hat{L}}$, which gives the localization probability as $ P_L(\alpha) \!=\! 2^{-N} \left( 1 + (-1)^N \cos(\alpha) \right).$ 
Numerically calculated variation of $P_L$ with the phase of the NOON state $\alpha$ is shown in  
Fig.~\ref{Fig_S_localization_oscillation}(c) for $N\!=\!2, \, 3$ and $4$.


\begin{figure}[]
\centering
\includegraphics[width=0.9\linewidth]{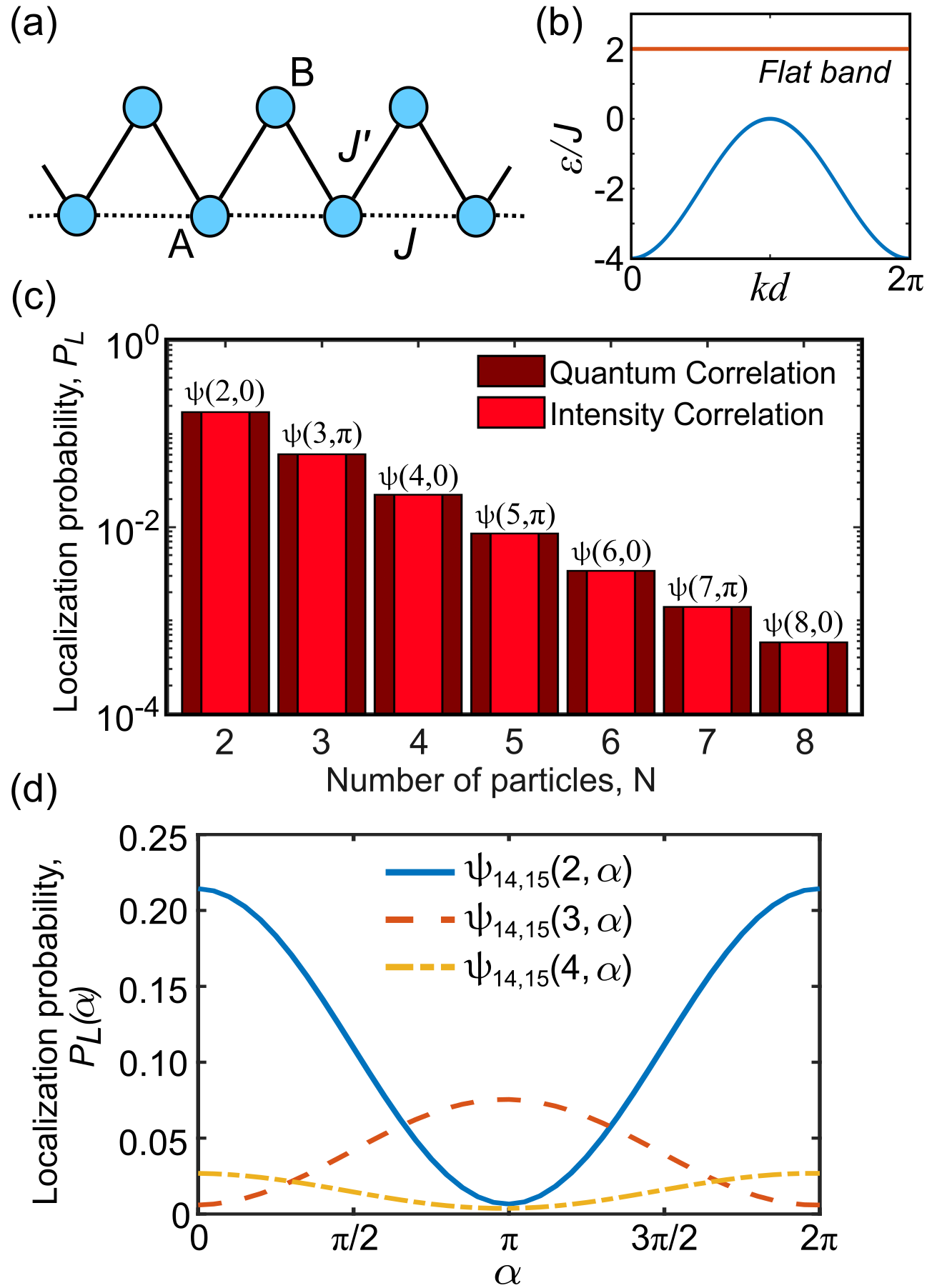} 
\caption{(a) Sketch of a sawtooth lattice. (b) Band structure for $J'\!=\!\sqrt{2}J$, showing an upper flat-band and a lower dispersive band. (c) Localization probability $P_L$ of all NOON state photons in the flat-band as a function of $N$.
The initial states are indicated above each bar. The brown (red) bars are obtained from quantum (intensity) correlation calculations. (d) Localization probability as a function of $\alpha$ shows maxima at $\alpha\!=\!0$ for $N\!=\!2$, $N\!=\!4$ NOON state ,and at $\alpha\!=\!\pi$ for $N\!=\!3$ case.
}
\phantomsection
\label{Fig_S_sawtooth}
\end{figure}

\section{NOON state in the flat-band sawtooth lattice}
\label{secS_sawtooth_lattice}

So far, we have discussed the dynamics of NOON states and localization-delocalization phenomena in a photonic rhombic lattice. In this section, we shall find out whether these phenomena can appear in other flat-band lattices. As an example, we consider a sawtooth lattice, consisting of two sites (A and B) per unit cell, see Fig.~\ref{Fig_S_sawtooth}(a). In this case, the tight-binding Hamiltonian is given by
\begin{align}
\hat{H} \!=\! \sum_{j} -J \hat{a}_{j+1}^\dagger \hat{a}_j 
- J' ( \hat{b}_j^\dagger \hat{a}_j + \hat{b}_j^\dagger \hat{a}_{j+1} ) + \text{H.c.} \, ,
\label{Eqn_S25}
\end{align}
where $\hat{a}_j$ is the bosonic creation operator on site $j$. The coupling between A sites is denoted by $J$, and that between A and B sites is $J'$. When $J'$ is tuned to $\sqrt{2}J$, the upper band becomes perfectly flat with eigenvalue $\varepsilon_{0}(k)\!=\!2J$, see Fig.~\ref{Fig_S_sawtooth}(b).  
Both flat-band and dispersive band eigenstates are $k$-dependent in this case.

For the rhombic lattice, the compact localized states are confined to the B and C sites of a unit cell. In contrast, the CLS in a sawtooth lattice lives on three sites, spanning over two unit cells. In this case, each CLS is non-orthogonal to its two nearest neighbors. These properties make a sawtooth lattice {\it fairly} different from the rhombic lattice.

In the sawtooth lattice, the CLS occupies three sites: $(A_n, B_n, A_{n+1})\!=\!(+1,-1,+1)/\sqrt{3}$. The phase and intensity-dependent localization-delocalization occurs when the NOON states are coupled to the sites with out-of-phase CLS amplitudes. Therefore, to explore the dynamics of the NOON states in the sawtooth lattice, we consider coupling the states at the A and B sites of a unit cell. 
The probability $P_L$ of finding all NOON state photons in the flat band is shown in Fig.~\ref{Fig_S_sawtooth}(c) as a function of $N$, alternately considering $\alpha\!=\!0$ and $\pi$. Here, the scaling is $P_L\!=\!X \cdot Y^{-N}$, where $X\!=\!1.5967$ and $Y\!=\!2.755$.
As in the case of the rhombic lattice in Fig.~\ref{Fig_3}
, for an even (odd)  $N$, $P_L$ is maximum at $\alpha\!=\!0$ $(\pi)$. However, it approaches to {\it nearly zero values} for the opposite phases, $\alpha\!=\!\pi$ $(0)$, see Fig.~\ref{Fig_S_sawtooth}(d).


%

\newpage

\end{document}